\documentclass[twocolumn, twocolappendix tighten, astrosymb, dvipsnames]{aastex631}
\usepackage{xspace}
\usepackage{graphicx}
\usepackage{amssymb}
\usepackage{amsmath}
\usepackage{float}
\usepackage{multirow}
\usepackage{textcomp}
\usepackage{gensymb}
\usepackage{hyperref}
\usepackage{natbib}
\usepackage{comment}
\usepackage{systeme}
\usepackage[Symbol]{upgreek}
\usepackage{xcolor}

\newcommand{\chandra}{\textit{Chandra}\xspace}
\newcommand{\xmm}{\textit{XMM-Newton}\xspace}
\newcommand{\lofar}{{LOFAR}\xspace}

\shorttitle{Diffuse radio emission in a low-mass cluster}
\shortauthors{Rajpurohit et al.}

\graphicspath{{./}{figures/}}

\begin{document}

\title{PSZ2\,G181.06+48.47 II: radio analysis of a low-mass cluster with exceptionally-distant radio relics}

\correspondingauthor{Kamlesh Rajpurohit}
\email{kamlesh.rajpurohit@cfa.harvard.edu}

\author[0000-0001-7509-2972]{Kamlesh Rajpurohit}
\affiliation{Center for Astrophysics \text{\textbar} Harvard \& Smithsonian, 60 Garden St., Cambridge, MA 02138, USA}

\author[0000-0001-8322-4162]{Andra Stroe}
\affiliation{Center for Astrophysics \text{\textbar} Harvard \& Smithsonian, 60 Garden St., Cambridge, MA 02138, USA}
\affiliation{Space Telescope Science Institute, 3700 San Martin Drive, Baltimore, MD 21218, USA}

\author[0000-0002-5671-6900]{Ewan O'Sullivan}
\affiliation{Center for Astrophysics \text{\textbar} Harvard \& Smithsonian, 60 Garden St., Cambridge, MA 02138, USA}

\author[0009-0009-4676-7868]{Eunmo Ahn}
\affiliation{Department of Astronomy, Yonsei University, 50 Yonsei-ro, Seodaemun-gu, Seoul 03722, Republic of Korea}

\author[0000-0002-1566-5094]{Wonki Lee}
\affiliation{Department of Astronomy, Yonsei University, 50 Yonsei-ro, Seodaemun-gu, Seoul 03722, Republic of Korea}

\author[0000-0001-5966-5072]{Hyejeon Cho}
\affiliation{Department of Astronomy, Yonsei University, 50 Yonsei-ro, Seodaemun-gu, Seoul 03722, Republic of Korea}
\affiliation{Center for Galaxy Evolution Research, Yonsei University, 50 Yonsei-ro, Seodaemun-gu, Seoul 03722, Republic of Korea}

\author[0000-0002-5751-3697]{M. James Jee}
\affiliation{Department of Astronomy, Yonsei University, 50 Yonsei-ro, Seodaemun-gu, Seoul 03722, Republic of Korea}
\affiliation{Department of Physics and Astronomy, University of California, Davis, One Shields Avenue, Davis, CA 95616, USA}

\author[0000-0002-0587-1660]{Reinout van Weeren}
\affiliation{Leiden Observatory, Leiden University, PO Box 9513, 2300 RA Leiden, The Netherlands}

\author[0000-0002-3754-2415]{Lorenzo Lovisari}
\affiliation{INAF, Istituto di Astrofisica Spaziale e Fisica Cosmica di Milano, via A. Corti 12, 20133 Milano, Italy}
\affiliation{Center for Astrophysics \text{\textbar} Harvard \& Smithsonian, 60 Garden St., Cambridge, MA 02138, USA}

\author[0000-0002-4462-0709]{Kyle Finner}
\affiliation{IPAC, California Institute of Technology, 1200 E California Blvd., Pasadena, CA 91125, USA}

\author[0000-0002-9714-3862]{Aurora Simionescu}
\affiliation{SRON, Netherlands Institute for Space Research, Sorbonnelaan 2, 3584 CA Utrecht, The Netherlands}

\author[0000-0002-9478-1682]{William Forman}
\affiliation{Center for Astrophysics \text{\textbar} Harvard \& Smithsonian, 60 Garden St., Cambridge, MA 02138, USA}

\author[0000-0001-5648-9069]{Timothy Shimwell}
\affiliation{ASTRON, the Netherlands Institute for Radio Astronomy, Postbus 2, NL-7990 AA Dwingeloo, the Netherlands}
\affiliation{Leiden Observatory, Leiden University, P.O. Box 9513, NL-2300 RA Leiden, the Netherlands}

\author[0000-0003-2206-4243]{Christine Jones}
\affiliation{Center for Astrophysics \text{\textbar} Harvard \& Smithsonian, 60 Garden St., Cambridge, MA 02138, USA}

\author[0000-0001-8812-8284]{Zhenlin Zhu}
\affiliation{SRON Netherlands Institute for Space Research, Niels Bohrweg 4, 2333 CA Leiden, The Netherlands}
\affiliation{Center for Astrophysics \text{\textbar} Harvard \& Smithsonian, 60 Garden St., Cambridge, MA 02138, USA}
\affiliation{Leiden Observatory, Leiden University, Niels Bohrweg 2, 2300 RA Leiden, The Netherlands}

\author[0000-0002-3984-4337]{Scott Randall}
\affiliation{Center for Astrophysics \text{\textbar} Harvard \& Smithsonian, 60 Garden St., Cambridge, MA 02138, USA}

\begin{abstract}
We report upgraded  Giant Metrewave Radio Telescope and Karl J. Jansky Very Large Array radio observations of a low-mass merging galaxy cluster PSZ2\,G181.06+48.47. This exceptional galaxy cluster hosts two megaparsec-scale diffuse sources, symmetrically located with respect to the cluster center and separated by about 2.6\,Mpc in projection. We detect these low surface brightness sources in our new high-frequency observations ($0.3{-}2$\,GHz) and classify them as radio relics associated with merger-driven shock fronts. The southwest relic exhibits an inverted morphology and shows evidence of spectral steepening in the post-shock region, potentially tracing a high Mach number shock ($\sim 4$) under the framework of diffusive shock acceleration. The northeast relic is found to be highly polarized with a 22\% average polarization fraction at 1.5\,GHz and aligned magnetic field vectors. Its spectral and polarization properties, along with the presence of a nearby tailed galaxy, support re-acceleration scenarios. The merger axis defined by the two relics is tilted by $\sim 45\degree$ with respect to the plane of the sky, which implies an unprecedented  physical separation of $\sim 3.5\,\rm Mpc$. We also detect a possible faint radio halo, suggesting weak turbulence in the central cluster region. We conclude that the faint double relics can be best explained by two outward moving shock waves in which particles are (re-)accelerated and that the cluster is in an evolved merger state. PSZ2\,G181.06+48.47 presents a unique opportunity to investigate particle acceleration in low mass systems characterized by large relic separations. 
\end{abstract}

\vspace{-3em}

\section{Introduction}
\label{sec:intro}

Galaxy clusters grow through accretion and via mergers of smaller galaxy clusters and groups. During mergers, large-scale shocks are driven into the intracluster medium (ICM), forming radio relics at cluster outskirts \citep[e.g.,][for a theoretical and observational review]{Brunetti2014, vanWeeren2019}. The detection of X-ray discontinuities/edges at the location of the relics confirms the connection between radio relics and ICM shocks \citep{Markevitch1997, Sarazin2013, Botteon2016, Shimwell2015}. Relics have typically steep radio spectra\footnote{We define the spectral index, $\alpha$, so that $S_{\nu}\propto\nu^{\alpha}$, where $S$ is the flux density at frequency $\nu$} ($\alpha\leq-1$ ) and are among the strongest linearly polarized sources in the sky, reaching polarization fractions as high as 60\% ($>1$\,GHz), with magnetic field orientations well aligned with the shock surface \citep[e.g.,][]{vanWeerenSci, Bonafede2012, Owen2014, Gennaro2021, Rajpurohit2022a, Wittor2019, Paola2021b}. 

Radio relic sizes, morphology and separations from the cluster center vary significantly, as does their degree of association with shocks detected via X-ray observations. This makes it challenging to obtain a self-consistent theoretical model of the origin of relics. According to the most widely accepted scenario \citep[e.g.][]{Brunetti2014, vanWeeren2019, Paul2023}, radio relics originate when particles at the shock front are accelerated via diffusive shock acceleration (DSA) mechanisms \citep{Drury1983, Hoeft2007}. However, the acceleration efficiency of weak ICM shocks ($\mathcal{M}\leq2.5$) appears too low to explain the high luminosity of many relics, if the particles are accelerated directly from the thermal ICM \citep{Botteon2020}. Alternative models include shock re-acceleration, where shocks re-accelerate fossil electrons previously injected by AGN \citep{Stroe2014, Stroe2016, Kang2011, Kang2016a, Kang2016b} or multiple shock scenarios \citep[MSS, e.g.][]{Kang2021, Inchingolo2022}. Both models help reconcile the lower acceleration efficiency with the high radio luminosity for weak-shock relics \citep{Bonafede2014, vanWeeren2017a, Gennaro2018}.

Mergers between galaxy clusters can also produce another type of steep spectrum diffuse radio source, known as a radio halo, which fills the central regions of clusters and typically follows the thermal ICM morphology \citep{Giovannini2004, Bonafede2013, Bonafede2022, Botteon2022a, Rajpurohit2021c}. Halos are believed to be produced by re-acceleration of (primary or secondary) particles via stochastic Fermi II-type processes driven by cluster-merger-induced ICM turbulence \citep{Brunetti2001, Brunetti2007, Petrosian2001}. Radio halos are also characterized by their steep spectral indices, namely $\alpha \leq -1.1$, and are usually unpolarized \citep[e.g., ][]{vanWeeren2012a, Hoang2018, Wilber2018, Botteon2020, Luca2021, Rajpurohit2021c, Bonafede2022, Sikhosana2023}. The radio power (at 1.4\,GHz and 144\,MHz) of halos is found to increase with the host cluster mass, therefore it is challenging to detect faint radio halos in low mass clusters. Only a handful of radio halos detected in clusters with mass $M_{500}\leq 5\times10^{14}M_{\odot}$ \citep[e.g.,][]{Knowles2016, Paul2017, Paul2021, Hoang2021, Botteon2021}.

\begin{figure*}[!thbp]
    \centering
    \includegraphics[width=0.95\textwidth]{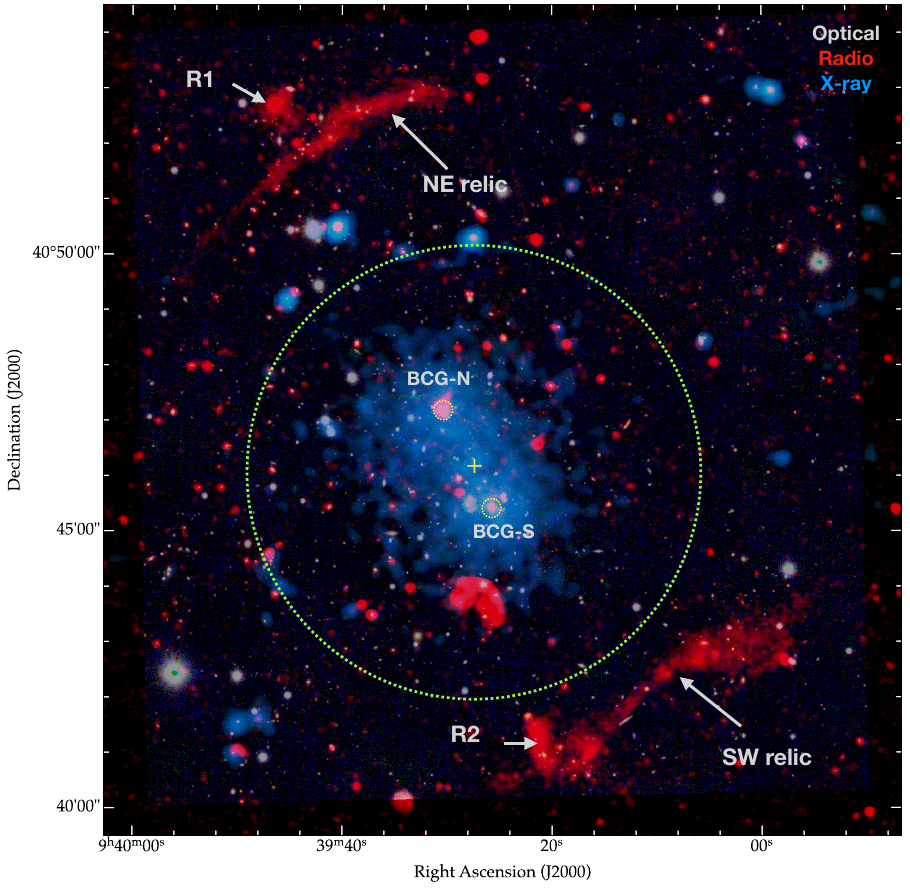}
    \vspace{-0.2cm}
 \caption{Radio, X-ray, and optical overlay of PSZ2\,G181.06+48.47, a post-core passage merging cluster system. The radio emission from \lofar \citep[144\,MHz,][]{Botteon2022a} and uGMRT Band\,3 (380\,MHz) is shown in the red color, while the blue traces the 0.5--2.0\,keV \xmm X-ray emission \citep{PSZ_Xray}. In the background, the color composite RGB optical image was created using PanSTARRS \textit{g}, \textit{r} and \textit{i} data. The yellow plus marks the center of the system. We label the most prominent diffuse radio sources (R2 is a radio galaxy, R1 is a diffuse source) and mark the two BCGs, strong radio point sources, with dotted circles. The green circle represents $R_{500}$ of the cluster.}
      \label{fig:overlay}
\end{figure*}

Radio relics are relatively rare, with about 10\% of Planck galaxy clusters hosting at least one relic \citep[e.g.][]{Jones2023}. Some merging clusters host double radio relics, i.e., relics located symmetrically with respect to the cluster center \citep[e.g.][]{Bonafede2014, DeGasperin2014, vanWeeren2019, Jones2023}, of which fewer than $30$ confirmed examples are known to date \citep{PSZ_Xray}. There are a handful of clusters with double relics that also host a radio halo, for example CIZA\,J2242.8+5301 \citep{vanWeerenSci, Stroe2013, Hoang2017, Gennaro2018}, PLCKG287.0+32.9 \citep{Bonafede2014,Stuardi2022}, PSZ1\,G108.18-11.53 \citep{deGasperin2015}, RXC\,J1314.4-2515 \citep{Stuardi2019, Knowles2022}. The detection rate of radio relics is impacted by the detailed physics of relic formation, compounded with the sensitivity constraints relevant at low surface brightness. Producing double relics requires very particular physical conditions. After the dark matter (DM) core passage in a binary, head-on collision of two clusters, two merger-induced shocks propagate in opposite directions along the merger axis \citep[e.g.][]{Skillman2011, vanWeeren2011, Lee2024}. However, the relics will only `light-up' in the radio at relatively large cluster-centric distances. The radio emission peaks $1-1.5$\,Gyr after core passage \citep[e.g.][]{2016MNRAS.462.2014D}: this is governed by the kinetic energy dissipated into shocks, which peaks at about 0.5\,$R_{\rm vir}$, and the cosmic ray (CR) production, which peaks at about 1\,Gyr after core passage \citep{Vazza2012, Ha2018}. When multiwavelength observation data is available, the merger scenario of double-relic clusters can be well constrained by piecing together information on the gas distribution, location of the shocks, etc. \citep[e.g.][]{Roettiger1999, Dawson2013, Ng2015, Lee2020, Kim2021, Cho2022, Albuquerque2024}. Therefore, double-radio relic clusters offer the unique opportunity to pose tight constraints on particle acceleration mechanisms within the context of a well understood merger scenario. Comparing clusters with double relics and with/without halos provides the ideal testbed for constraining the cluster properties that govern the formation of radio halos during mergers \citep{Bonafede2017, Jee2016, Golovich2019, Zhang2020}.

Although only a few more than two dozen double relics are currently known, there could potentially be many more awaiting discovery in the era of large radio surveys \citep[e.g.][]{Nuza2017}. Instruments such as the LOw-Frequency ARray \citep[\lofar,][]{Haarlem2013}, the upgraded Giant Metrewave Radio Telescope \citep[uGMRT,][]{Gupta2017}, and Square Kilometer Array precursors such as MeerKAT \citep{MeerKAT2016} with their combination of $5^{\prime\prime}$ resolution, high sensitivity, and mid and low radio frequency capabilities uniquely suited to detecting steep spectrum radio sources, coupled with a large field-of-view, are already changing the way we study diffuse radio sources in clusters. With these instruments, complex, diffuse radio sources are being discovered in a new parameter space, covering lower masses and higher redshifts than was possible before with shallower surveys at lower resolution \citep[e.g.][]{Parekh2020, Gennaro2021, Botteon2022a, Lee2022, Knowles2022, Sikhosana2024, Chatterjee2024}.

\section{PSZ2\,G181.06+48.47}
A spectacular example of a double-relic system is PSZ2\,G181.06+48.47 and its associated Mpc-wide diffuse sources (Figure~\ref{fig:overlay}). The cluster was originally discovered through the red sequence method in the Sloan Digital Sky Survey (SDSS) \citep{Koester2007} and was later confirmed through the redMaPPer algorithm \citep{Rykoff2014} and through its Sunyeav-Zel'dovich (SZ) effect by Planck \citep{Planck2016}. More recently, spectroscopy in the region of the source from Sloan Digital Sky Survey Data Release 18 \citep[SDSS DR18,][]{2023ApJS..267...44A} confirms the presence of an overdensity of galaxies at $z\sim0.234$ \citep{PSZ_Xray}. 

As a relatively low-mass cluster ($M_{\rm SZ}=4.2^{+0.5}_{-0.5}\times10^{14}$\,M$_{\odot}$, \citealp{Planck2016}, weak lensing $M_{\rm 500,WL}=2.9^{+0.75}_{-0.69}\times10^{14}$\,M$_{\odot}$, \citealp*{PSZ_WL}, X-ray $M_{\rm 500,X}=2.57^{+0.37}_{-0.38}\times10^{14}$\,M$_{\odot}$, \citealp*{PSZ_Xray}) at $z=0.234$, and seemingly devoid of any particularly interesting features, PSZ2\,G181.06+48.47 initially eluded further study. A recent X-ray study suggests that this cluster is one of the most disturbed clusters in the \textit{Planck} sample, exhibiting a complex morphological and thermodynamic structure \citep{PSZ_Xray}. 

The redshift and mass of the cluster and the sizes and luminosities of the diffuse sources place them within the reach of deep, pointed, state-of-the-art observations, but below the limits of classical radio survey detection limits. The twin sources were detected in the recent \lofar 144\,MHz observation thanks to the unique combination of observing frequency, sensitivity and resolution \citep[][]{Shimwell2019, Botteon2022a}. With otherwise typical radio powers and sizes, the two putative relics are located at exceptionally far ($>1.2$\,Mpc) away from the putative cluster core, beyond $R_{500}\sim0.93$\,Mpc close to $R_{200}\sim1.43$\,Mpc, given the low mass of the host cluster \citep{PSZ_WL, PSZ_Xray}. The bright diffuse sources located in a low-mass cluster at such large cluster-centric distances, where the plasma density is low, can place valuable tests on relic formation models. Are the relics associated with particularly strong shocks? Is the CR density higher than expected? Could a more complex cluster merger history explain the morphology and location of the two diffuse sources?

\vspace{-1.2em}
\begin{deluxetable*}{l c c c c c c c c}[!htbp]
\tablecaption{Overview of VLA, uGMRT, and LOFAR observations.}
\tablehead{\multirow{1}{*}{}& \multicolumn{3}{c}{VLA L-band} & \multicolumn{2}{c}{uGMRT}  &\multirow{1}{*}{LOFAR HBA$^{\ast}$}   \\  
 \cline{2-4}  \cline{5-6} 
& B configuration & C configuration & D configuration& Band\,4& Band\,3& &}
\startdata
Frequency range &   1--2\,GHz    &   1--2\,GHz    &   1--2\,GHz    &   550--850\,MHz &  300--500\,MHz    & 120--168\,MHz \\ 
Channel width   &   1\,MHz      &   1\,MHz & 1\,MHz & 48.8\,kHz & 130\,kHz &    12.2\,kHz \\ 
Correlations    &   full Stokes &   full Stokes &    full Stokes &    RR and LL   &  RR and LL   &  full Stokes  \\
On source time  &   4\,hrs       &  2\,hrs &  2\,hrs &   9\,hrs& 9\,hrs    & 8\,hrs  \\
\enddata
\tablecomments{VLA observations were recorded in 16 spectral windows (64 channels in each spectral window); $^{\ast}$Observation presented in \cite{Botteon2022a}.}
\label{tab:radio_obs}
\end{deluxetable*} 
\vspace{-1.2em}

To leverage the unique insights PSZ2\,G181.06+48.47 offers into particle acceleration in low mass clusters, we began a multiwavelength observational campaign to unveil the thermal and non-thermal cluster properties in the context of its merger history. In the present paper, we focus on unveiling the nature, physics and formation mechanisms of the diffuse radio sources associated with the cluster through broad-band radio observations spanning 120\,MHz to 1.5\,GHz. Through a detailed analysis of the thermodynamic properties of the ICM, \citet{PSZ_Xray} characterize the subclusters, identify merger-associated discontinuities and pose constraints on the merger history, while \citet{PSZ_WL} focuses on a weak lensing (WL) analysis of the cluster aimed at understanding the mass properties of the clusters, its merger history and, more broadly, constraining DM properties.

This paper is structured as follows. In Section~\ref{sec:obs}, we introduce the new VLA and uGMRT observations, present our data reduction strategy, and list ancillary data. Section~\ref{sec:analysis} describes the total intensity, spectral and polarization analysis techniques employed in exploring the data, while in Section~\ref{sec:results} we present the results of our analysis, including morphological, integrated and resolved spectral, color-color, and polarimetric properties of the sources. In Section~\ref{sec:discussion}, we investigate the origin of the diffuse radio sources in the context of the merger scenario that best fits PSZ2\,G181.06+48.47. Concluding remarks can be found in Section~\ref{sec:conclusions}. Throughout this paper, we adopt a flat $\Lambda$CDM cosmology with $H_{\rm{ 0}}=69.6$\,km\,s$^{-1}$\,Mpc$^{-1}$, $\Omega_{\rm{ m}}=0.286$, and $\Omega_{\Lambda}=0.714$ \citep{Wright2006}. At the cluster redshift of $z=0.234$, $1\arcsec$ corresponds to a physical scale of 3.753\,kpc. We use a cluster center defined to sit between the two subclusters at ${\rm RA}=144.866\degree$ and ${\rm Dec}=40.774\degree$ \citep{PSZ_Xray}.

\section{Observations and data reduction}
\label{sec:obs}

\subsection{Very Large Array}

The cluster was observed with the Karl G. Jansky Very Large Array (VLA) using the L-band (1-2\,GHz) in the B, C, and D configurations (project code: SL0429, PI: A. Stroe, see Table~\ref{tab:radio_obs}). For each configuration, 3C\,286 was included as a flux calibrator, observed for 6 minutes at the start of the observation. One of two radio sources (J0832+4913 or J1006+3454) was included as a phase calibrator.

The data were calibrated and imaged with the Common Astronomy Software Applications \citep[\texttt{CASA};][]{McMullin2007,casa2022} package. Data obtained from different observing runs were calibrated separately but in the same manner. The data were Hanning smoothed, inspected for radio frequency interference (RFI), and the affected data were flagged using the \texttt{tfcrop} mode from the \texttt{flagdata} task. Low-amplitude RFI was flagged using {\tt AOFlagger} \citep{Offringa2010}. Following flagging, we determined and applied elevation-dependent gain tables and antenna offset positions. To prevent the flagging of good data due to the bandpass roll-off at the edges of the spectral windows, we corrected for the bandpass using the calibrator 3C\,48. We used the L-band 3C\,137 and 3C\,286 models provided by the {\tt CASA} software package and set the flux density scale according to \cite{Perley2013}. An initial phase calibration was performed using both calibrators over a few channels per spectral window. We then corrected the antenna delays and determined the bandpass response using the calibrator 3C\,147. Applying the bandpass and delay solutions, we proceeded with the gain calibration.

\vspace{-1.2em}
\begin{deluxetable*}{c c c r c c c r}[!thbp]
\tablecaption{Imaging properties of radio maps used in the analysis}
\tablehead{& Name & Restoring Beam & Robust  & \textit{uv}-cut & \textit{uv}-taper & RMS noise\\ 
&&& parameter &&&$\upmu\rm Jy\,beam^{-1}$}
\startdata
\multirow{3}{3cm}{LOFAR HBA (120--169\,MHz)} &IM1&$8.1\arcsec \times 5.3\arcsec$&$-0.5$&$-$&$-$&70\\
  &IM2&$20\arcsec \times 20\arcsec$&$-0.5$&$-$&15\arcsec&200\\
  &IM3&$20\arcsec \times 20\arcsec$&$-0.5$&$  \geq\rm0.15\,k\uplambda$&15\arcsec&210\\
\hline
 \multirow{4}{3cm}{uGMRT Band\,3 (300--500\,MHz)} &IM4&$6.8\arcsec \times 5.6\arcsec$&0.0&$-$&$-$&15\\
  &IM5&$20\arcsec \times 20\arcsec$&$0.0$&$ -$&12\arcsec&42\\
  &IM6&$20\arcsec \times 20\arcsec$&$-0.5$&$ \geq\rm0.15\,k\uplambda$&15\arcsec&65\\ 
  &IM7&$25\arcsec \times 25\arcsec$&$0.0$&$ -$&25\arcsec&90\\
 \hline   
\multirow{3}{3cm}{uGMRT Band\,4 (550--850\,MHz)} &IM8&$5\arcsec \times 5\arcsec$&0.0&$-$&$-$&10\\
  &IM9&$20\arcsec \times 20\arcsec$&$0.0$&$ -$&12\arcsec&22\\
  &IM10&$20\arcsec \times 20\arcsec$&$-0.5$&$ \geq\rm0.15\,k\uplambda$&15\arcsec&25\\
\hline   
\multirow{3}{3cm}{VLA L-band (1--2\,GHz)} &IM11&$8.5\arcsec \times 7.4\arcsec$&0.0&$-$&$-$&9\\
  &IM11&$20\arcsec \times 20\arcsec$&$0.0$&$-$&15\arcsec& 14\\
  &IM13&$20\arcsec \times 20\arcsec$&$-0.5$&$ \geq\rm0.15\,k\uplambda$&15\arcsec&17\\
\enddata
\tablecomments{Imaging was performed in \texttt{WSCLEAN} using {\tt multiscale} and with the {\tt Briggs} weighting scheme.}
\label{tab:imaging}
\end{deluxetable*}
\vspace{-1.2em}

Following initial calibration and flagging, several rounds of self-calibration were performed to refine the calibration for each individual data set. We imaged the data in CASA with the W-projection algorithm \citep{Cornwell2008}. Clean masks made using the {\tt PyBDSF} source detection package \citep{Mohan2015} were employed for each imaging step. The spectral index and curvature were taken into account during deconvolution using nterms\,=\,2 \citep{Rau2011}. After self-calibration, the B, C, and D configurations data were combined, achieving denser \textit{uv} coverage. Final imaging of the combined data sets (BCD configurations) was completed in {\tt WSClean} \citep{Offringa2014}, using Briggs weighting, and employing the wideband and multi-scale algorithms \citep{Offinga2017}. The resulting images, providing an optimal balance between resolution and noise levels, were corrected for the primary beam attenuation in WSClean. The detailed parameters for the images are listed in Table~\ref{tab:imaging}.

\subsection{uGMRT}

PSZ2\,G181.06+48.47 was observed with the upgraded Giant Metrewave Radio Telescope (uGMRT) in Band\,4 and Band\,3 (project code 38\_062, PI: A. Stroe) using the GMRT Wideband Backend (GWB). The observations were conducted in multiple observing runs between September 2020 and January 2021, with 3C\,147 and 3C\,286 included as flux calibrators. We summarize the observational details in Table~\ref{tab:radio_obs}.

The uGMRT data were processed using Source Peeling and Atmospheric Modeling \citep[SPAM;][]{Intema2009}, which includes both direction-independent and direction-dependent calibration. The flux densities of the primary calibrator were set according to \cite{Scaife2012}. Following flux density scale calibration, the data were averaged, flagged, and corrected for the bandpass. The phase gains were corrected with a global sky model obtained with the GMRT narrowband data. The calibrated subbands were combined to produce deep, full continuum images. The final deconvolution was performed in WSClean using multiscale and Briggs weighting with robust parameter 0 (see Table~\ref{tab:imaging}) and corrected for primary beam.

\subsection{LOFAR}

The cluster was observed with LOFAR High Band Antenna (HBA) dual inner mode as part of the LOFAR Two-metre Sky Survey. Our analysis uses the calibrated data and the images released as part of the LOFAR DR2 \citep[see Table~\ref{tab:imaging} and][]{Shimwell2019}. For a detailed description of the observation and data reduction, we refer to \cite{Botteon2022a}.

\subsection{Ancillary observations}

To aid in the interpretation of the radio data, we also employ multiwavelength observations compiled through the Vizier database \citep{2000A&AS..143...23O}, including optical spectroscopy from the SDSS DR 18 \citep{2023ApJS..267...44A}. We also make use of data from the first data release from Panoramic Survey Telescope and Rapid Response System \citep[Pan-STARRS,][]{2020ApJS..251....7F} and new, targeted \chandra and \xmm X-ray observations \citep{PSZ_Xray}.

\section{Spectral and polarization analysis techniques}
\label{sec:analysis}

We describe below the radio, spectral index, and polarization maps we produced to investigate the morphological, spectral, and polarimetric properties of the radio sources in PSZ2\,G181.06+48.47.

\subsection{Radio Stokes I maps}

\begin{figure*}[!thbp]
    \centering
    \includegraphics[width=0.97\textwidth]{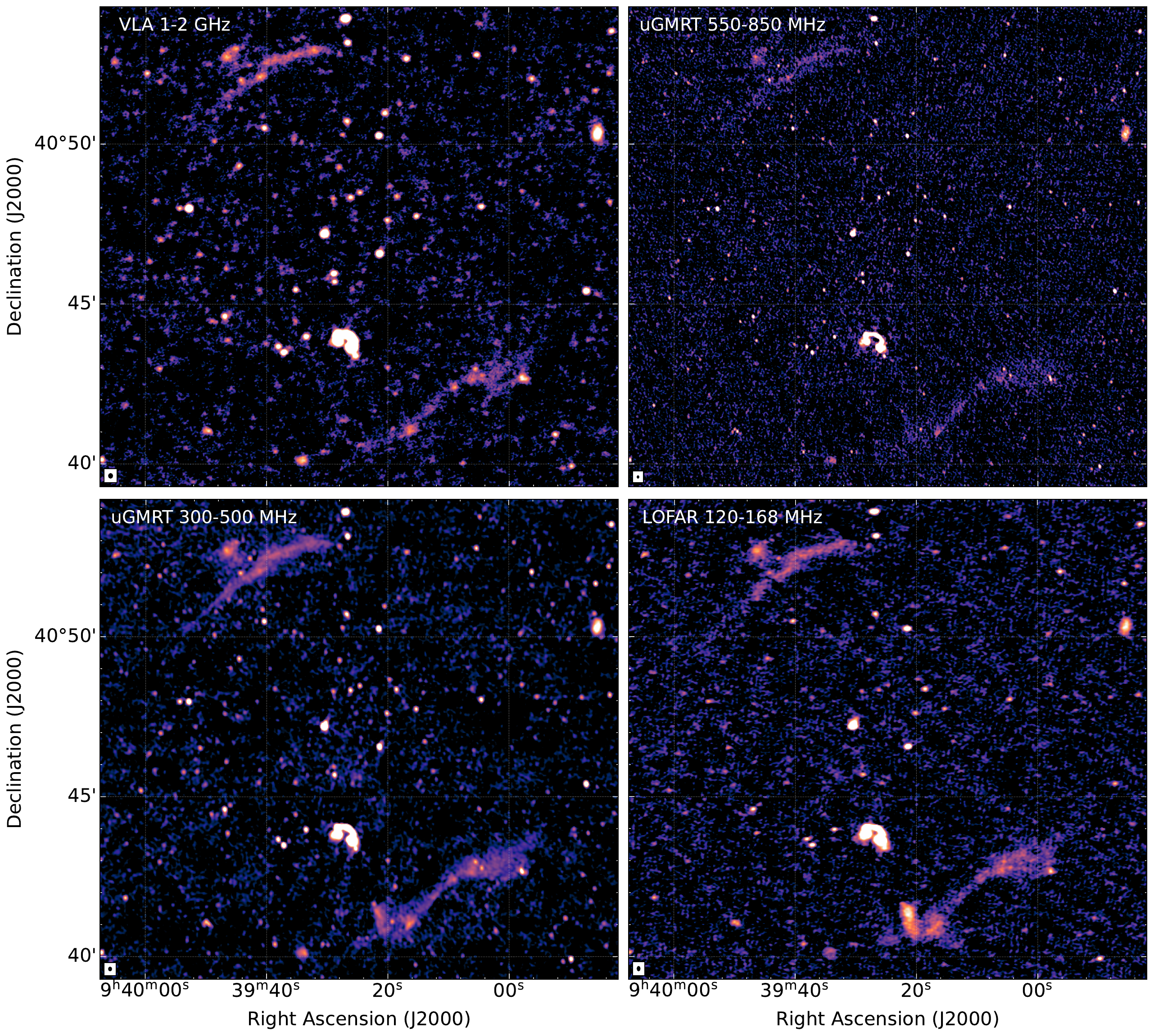}
    \caption{High-resolution VLA L-band (top left), uGMRT Band\,4 (top right), uGMRT Band\,3 (bottom left), and LOFAR HBA (bottom right) images of PSZ2\,G181.06+48.47 in square-root scale. The beam size is indicated in the bottom left corner of each image. Detailed image parameters are listed in Table~\ref{tab:imaging}.}
    \label{fig:high_res}
\end{figure*}

\begin{figure*}[!thbp]
    \centering
    \includegraphics[width=0.97\textwidth]{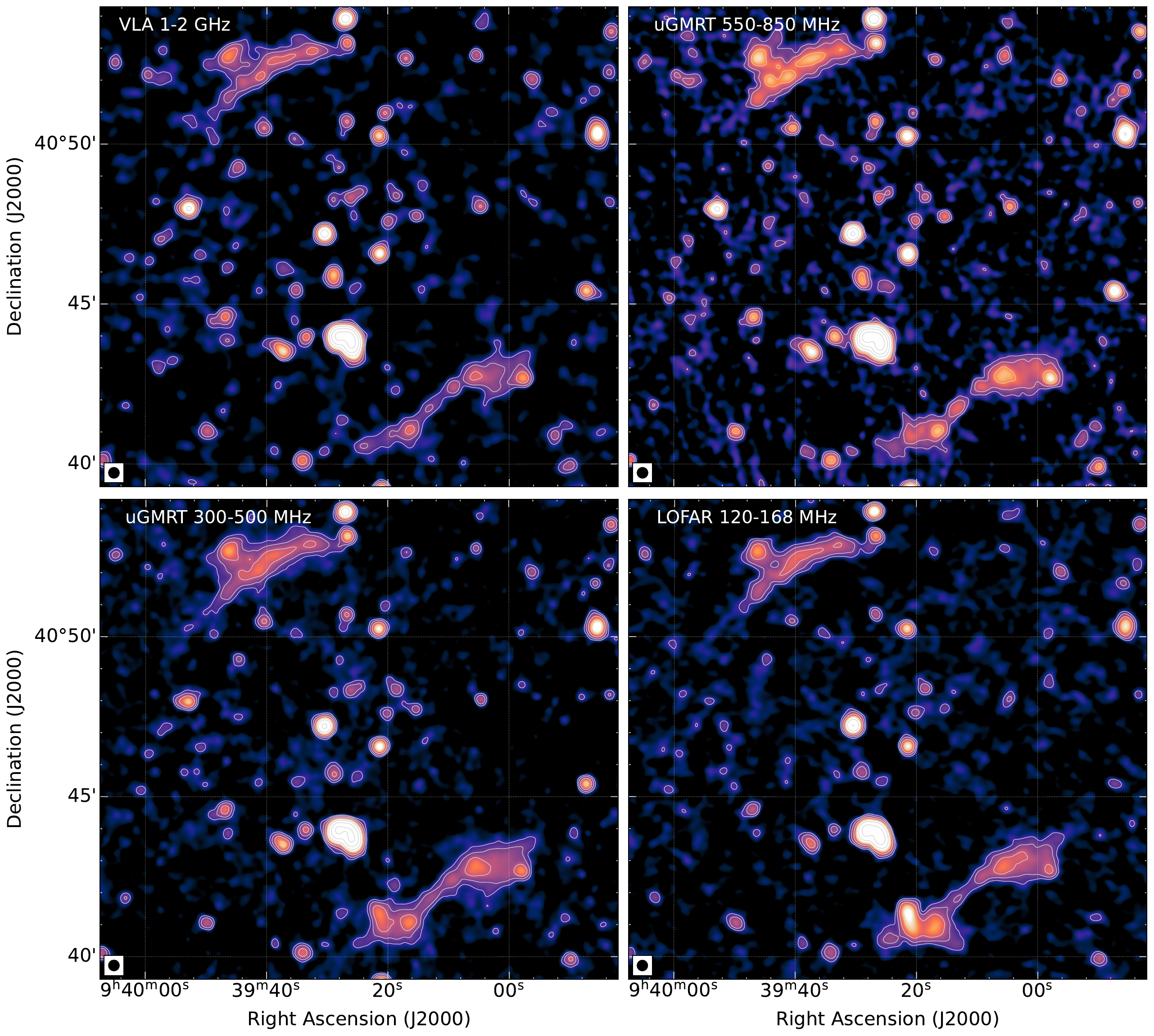}
    \caption{Low-resolution VLA L-band (top left), uGMRT Band\,4 (top right), uGMRT Band\,3 (bottom left), and LOFAR HBA (bottom right) images of the PSZ2\,G181.06+48.47 cluster. All images have a common resolution of 20{\arcsec}. The beam size is indicated in the bottom left corner of each image.}
    \label{fig:low_res}
\end{figure*}

We label the most prominent sources in Figure~\ref{fig:overlay}, while in Figure~\ref{fig:high_res}, we show high-resolution radio images of the cluster. 
To highlight radio emission on a range of spatial scales of interest, we also prepared a set of medium resolution, 20{\arcsec}-resolution images shown in Figure~\ref{fig:low_res}. These maps are created without any uv-cut. High-resolution zoom-in maps on the most prominent patches of diffuse emission are shown in the left panels of Figures~\ref{fig:NE_R1_overlay} and~\ref{fig:SW_R2_overlay}. For image properties, see Table\,\ref{tab:imaging}.

Additionally, to identify the lowest surface brightness radio features in the cluster, we also created our lowest resolution (25\arcsec resolution) uGMRT Band\,3 image using $\texttt{robust}=0.5$ and \textit{uv}-tapering (discussed in Section\,\ref{sec:halo}). 

\subsection{Spectral analysis}

To examine the spectral characteristics of the diffuse radio sources in the PSZ2\,G181.06+48.47 field across a broad frequency range, we combined our uGMRT Band\,3 (300--500\,MHz), Band\,4 (550--850\,MHz), and VLA (1--2\,GHz) observations with the published LOFAR HBA (120--169\,MHz) observations \citep{Botteon2022a}. To ensure consistency in the flux density distribution at all observed frequencies obtained from different radio interferometric telescopes, it is crucial that all images reflect structures over the same range of angular scales \citep[e.g.,][]{Stroe2016, Rajpurohit2020a}. Therefore, we imaged the radio data using a Briggs weighting scheme (${\tt roboust=-0.5}$), \textit{uv}-tapering, and a common lower \textit{uv}-cut at $0.15\rm k \lambda$, the shortest, well-sampled baseline of the uGMRT data. This \textit{uv}-cut was applied to the VLA L-band and LOFAR data. For spectral analysis, we created a set of images with different resolutions: 20\arcsec, 10\arcsec, and $7\arcsec$. The 20{\arcsec} resolution was chosen to properly exclude the contamination from unrelated sources, while allowing the recovery of diffuse flux from low surface brightness regions. The high-resolution maps (10{\arcsec} and $7\arcsec$) enable us to capture fine spectral index details and trends across both relics. 

The flux density scale of all observations (LOFAR, uGMRT, and VLA) was confirmed by comparing the spectra of compact sources between 144\,MHz and 1.5\,GHz. The uncertainty in the flux density measurements was estimated as:
\begin{equation}
\Delta S =  \sqrt {(f \cdot S)^{2}+{N}_{{\rm{ beams}}}\ (\sigma_{{\rm{rms}}})^{2}},
\end{equation}
where $f$ is the absolute flux density calibration uncertainty, $S$ is the flux density, $\sigma_{{\rm{ rms}}}$ is the RMS noise, and $N_{{\rm{beams}}}$ is the number of beams. We assumed absolute flux density uncertainties of 10\% for LOFAR HBA \citep{Shimwell2022}, uGMRT Band\,3, and uGMRT Band\,4 \citep{Chandra2004}, and 4\% for the VLA L-band data \citep{Perley2013}. 

The high resolution spectral index maps of the NE and SW relics, between 144\,MHz and 380\,MHz, are shown in the right panels of Figures~\ref{fig:NE_R1_overlay} and~\ref{fig:SW_R2_overlay}. Pixels below $3\sigma_{\rm rms}$ were blanked at both frequencies. For the purpose of measuring integrated spectral indices, we used the 20{\arcsec} resolution images. The specific regions from which flux densities were extracted are indicated in the right panel of Figure~\ref{fig:spectrum}, where point sources were masked.

\vspace{-1.2em}
\begin{deluxetable*}{*{10}{c}}[!thbp]
\tablecaption{Radio measurements for the diffuse radio sources in the cluster PSZ. See Figures~\ref{fig:spectrum} and~\ref{fig:RM_map} for the source labeling.}
\tablehead{\multirow{1}{*}{Source} &\multirow{1}{*}{LOFAR} &\multicolumn{2}{c}{uGMRT} &\multirow{1}{*}{VLA} &\multirow{1}{*}{$\rm LLS^{\dagger}$} & \multirow{1}{*}{$\alpha$}$^{\star}$ &\multirow{1}{*}{$P_{150\,\rm MHz}$} &\multirow{1}{*}{$P_{1.5\,\rm GHz}$}\\
 \cline{3-4} 
& $S_{\rm144\,MHz}$ &${S_{\rm 380\,MHz}}$&${S_{\rm 700\,MHz}}$ & ${S_{\rm1.5\,GHz}}$& &\\
  & (mJy) & (mJy) & (mJy) & (mJy) & (Mpc) & & $\rm 10^{24} W\,Hz^{-1}$& $\rm 10^{24} W\,Hz^{-1}$
}
\startdata
NE & $29.4\pm3.0$&$13.6\pm1.0$ & $6.8\pm0.7$& $3.5\pm0.5$ &$\sim1.3$&$-0.92\pm0.04$& $4.81\pm0.50$& $0.57\pm0.08$&\\
SW& $40\pm4.5$& $14.8\pm1.0$& $6.4\pm0.6$& $3.3\pm0.3$&$\sim1.2$& $-1.09\pm0.05$& $6.78\pm0.77$ & $0.56\pm0.05$ &\\
R1  & $9.4\pm1.5$ &$4.7\pm0.4$&$2.6\pm0.2$&$1.46\pm0.06$&$\sim0.25$& $-0.81\pm0.05$&-&-&\\
R2  & $15.0\pm2.0$& $3.2\pm0.3$& $1.07\pm0.05$&$0.33\pm0.02$&$\sim0.20$ &$-1.65\pm0.05$&-&-&\\
Halo  & $-$& $2.1\pm0.3$& $-$&$-$&$\sim1.0^{\ddagger}$ &$<-2$& $3$ & $-$ & \\
\enddata
\tablecomments{$^{\dagger}$The LLS measured from 144\,MHz map; $P$ is the radio power; $^{\star}$The integrated spectral index obtained by fitting a single power-law between 144\,MHz and 1.5\,GHz. $^{\ddagger}$Measured at 380\,MHz.}
\label{tab:flux}   
\end{deluxetable*} 
\vspace{-1.2em}

The medium resolution spatially resolved spectral index maps, created between 144\,MHz and 380\,MHz and between 144\,MHz and 1.5\,GHz, are shown in Figure~\ref{fig:index_maps}. Pixels were excluded if the flux density at both frequencies is below $3\sigma_{\rm rms}$.

\subsection{Color-color plots}
Radio color-color plots, in which the spectral index at low frequencies is plotted against the spectral index at high frequencies, are a powerful tool for analyzing the spectral properties of radio sources and for identifying overlapping regions \citep{Katz1993, KatzStone1997, vanWeeren2011a, Stroe2013, Rajpurohit2020a, Rajpurohit2022b}. The low-frequency spectral index is estimated between 144\,MHz and 380\,MHz, while the high frequency is between 700\,MHz and 1.5\,GHz. In Figure~\ref{fig:cc_plot}, we show the color-color plot obtained for the diffuse sources in PSZ2\,G181.06+48.47. Points that fall on the unity line ($\alpha^{380\rm\,MHz}_{144\rm\,MHz}$=$\alpha_{700\rm\,MHz}^{1.5\rm \,GHz}$) indicate a simple power-law spectrum, where the spectral index remains constant across frequencies. Any deviations from the power line imply negative (convex) or positive (concave) curvature. According to the spectral index convention we use, the curvature is negative for a convex spectrum.

\subsection{Polarization analysis}
\label{sec:polarization_analysis}
VLA L-band (1--2\,GHz) images of the Stokes I, Q, and U were obtained with \texttt{CASA}, employing ${\tt robust}= 0.5$ weighting and a \textit{uv} taper of 15{\arcsec}. The Stokes cubes affected by artifacts or exhibiting a high level of noise were excluded from the analysis. A total of 42 Stokes IQUV cubes were used for the whole 1--2\,GHz band. As the output Stokes cubes have slightly different resolutions, all Stokes cubes were convolved to a common restoring beam of 25{\arcsec}. Since the diffuse sources of interest exhibit low surface brightness, this resolution allows us to get a high signal-to-noise ratio and recover low surface brightness emission. Each image was corrected for the primary beam. 

We performed Rotation Measure (RM) Synthesis \citep[RM-synthesis:][]{Brentjens2005} on the Stokes IQUV cubes using the \texttt{pyrmsynth} {\footnote{\url{https://github.com/mrbell/pyrmsynth}}} code. The RM cube was created between $\pm300\rm\,rad \,m^{-2}$ with a bin size of $\rm 5\,rad\,m^{-2}$. Since the first sidelobes of the RM spread function were significant, we cleaned down to five times the noise level of full-bandwidth Q and U images using the \texttt{RM-CLEAN} algorithm. We used a polarization intensity ($P$) image at a central frequency of 1.5\,GHz generated by the \texttt{pyrmsynth} code to create a fractional polarization map via:
\begin{equation}
 \Pi=P/I ,
\end{equation}
where I is the Stokes I image at 1.5\,GHz. The polarization angle is then calculated using 
\begin{equation}
\chi=0.5 \arctan  \left( \frac{U}{Q}\right).
\end{equation}

\begin{figure*}[!thbp]
    \centering
    \includegraphics[width=0.95\textwidth]{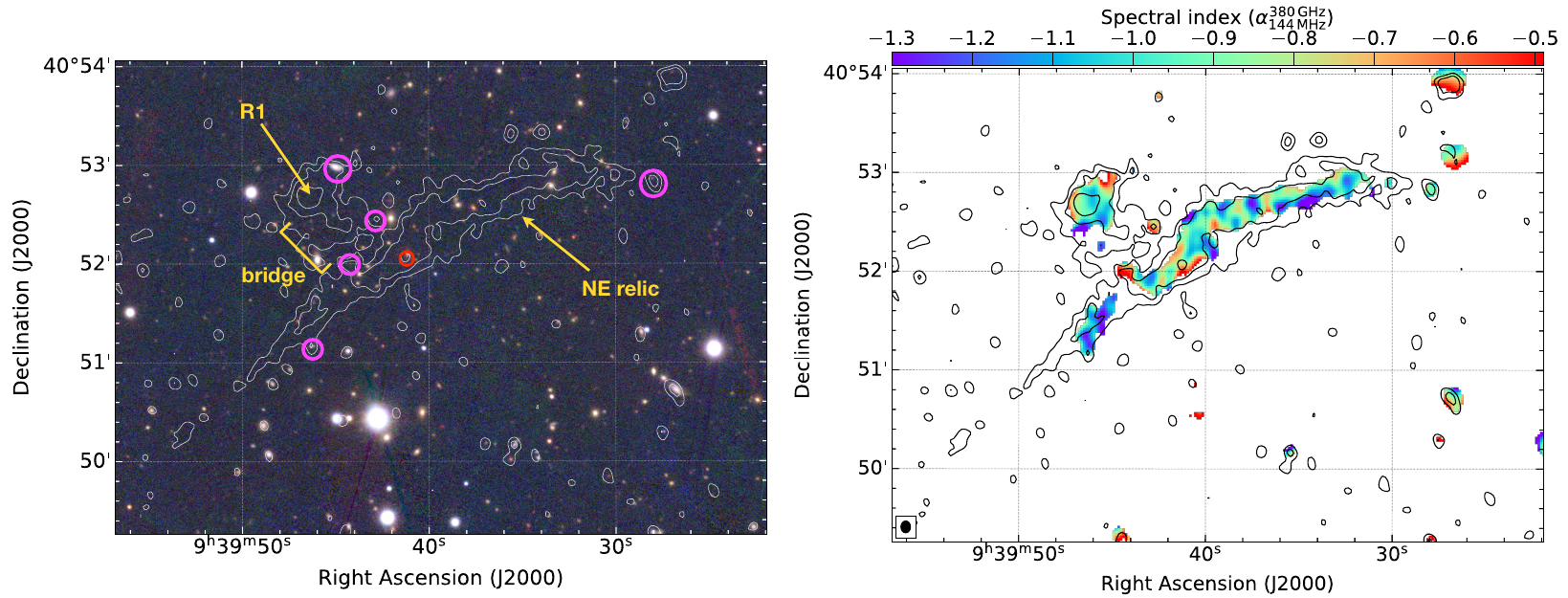}
    \caption{\textit{Left:} PanSTARRS image overlaid with uGMRT Band\,4 high resolution ($7\arcsec \times 6\arcsec$) point-source subtracted image of the NE relic. The beam size is indicated in the bottom left corner of each image. Contours drawn at $[1, 2, 4, 8 ...]\times 3.0\sigma_{\rm rms}$. The discrete sources subtracted from the \textit{uv} data are marked with circles with only one having an available redshift (red circle). \textit{Right:} {10\arcsec}-resolution spectral index map (144\,MHz--380\,MHz) of the NE relic, with uGMRT Band\,3 contours drawn at [1, 2, 4, 8...] $\times 4\,\sigma_\textrm{rms}$.}
    \label{fig:NE_R1_overlay}
\end{figure*}

\begin{figure*}[!thbp]
    \centering
    \includegraphics[width=0.95\textwidth]{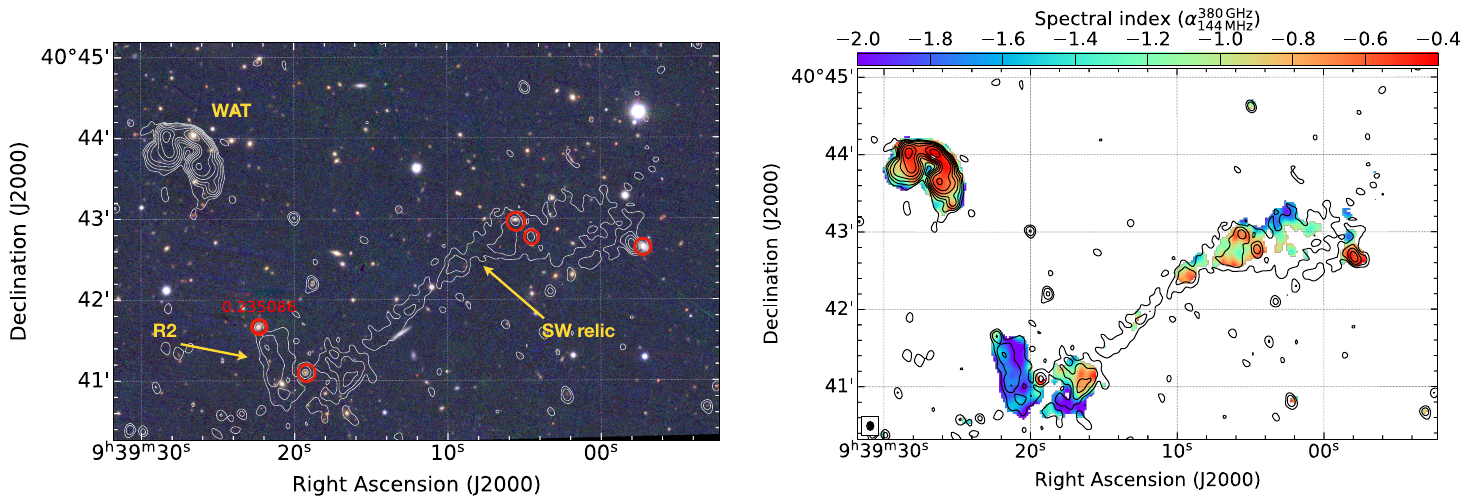}
    \caption{\textit{Left}: PanSTARRS zoom-in view of the SW relic overlaid with uGMRT Band\,3 {10\arcsec}-resolution radio contours. \textit{Right}: High-resolution ({10\arcsec}) spectral index map of the SW relic between 144\,MHz and 380\,MHz. uGMRT Band\,3 Stokes I contours are plotted at [1, 2, 4, 8, ...] $\times 3\,\sigma_\textrm{rms}$, where $\sigma_\textrm{rms}=18$\,$\mu$Jy\,beam$^{-1}$.}
    \label{fig:SW_R2_overlay}  
\end{figure*}

\begin{figure*}[!thbp]
    \centering
    \includegraphics[width=0.445\textwidth]{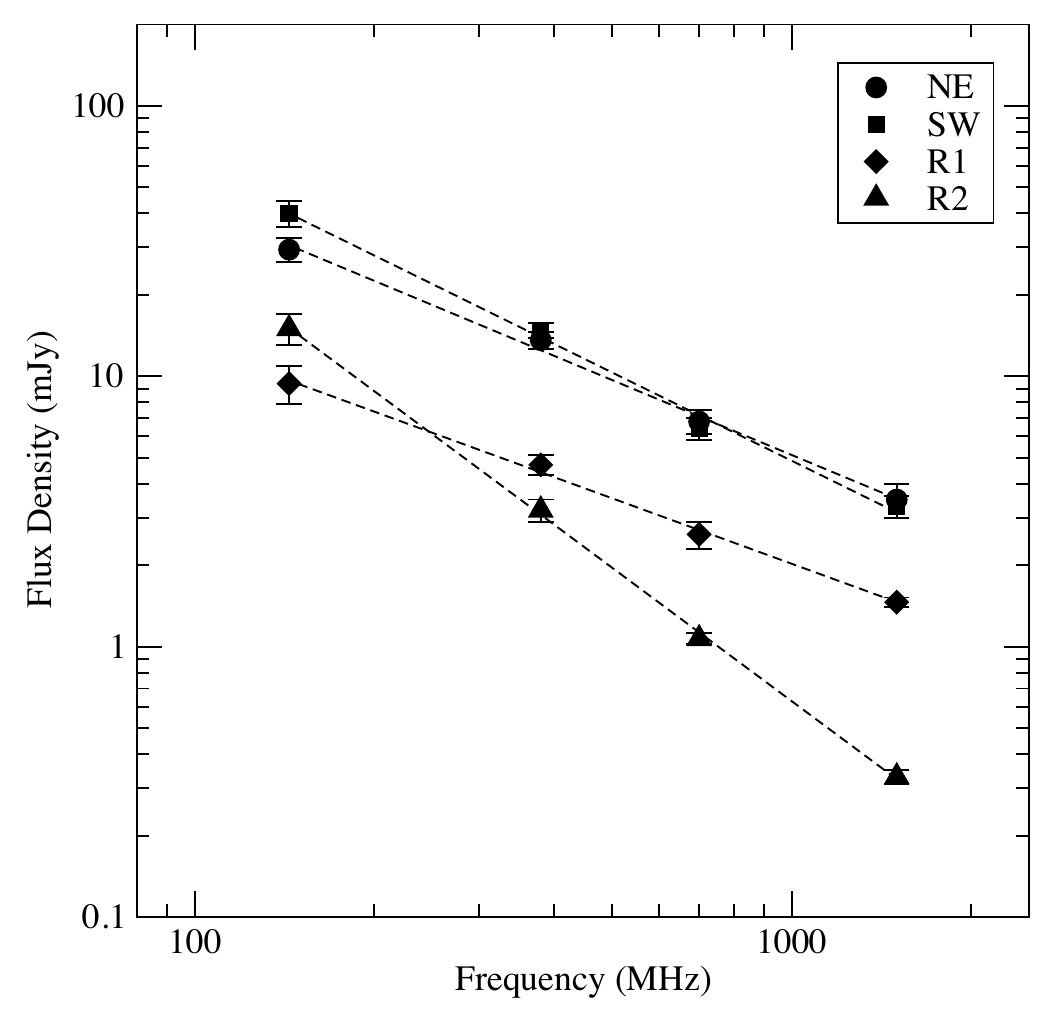}
    \includegraphics[width=0.49\textwidth]{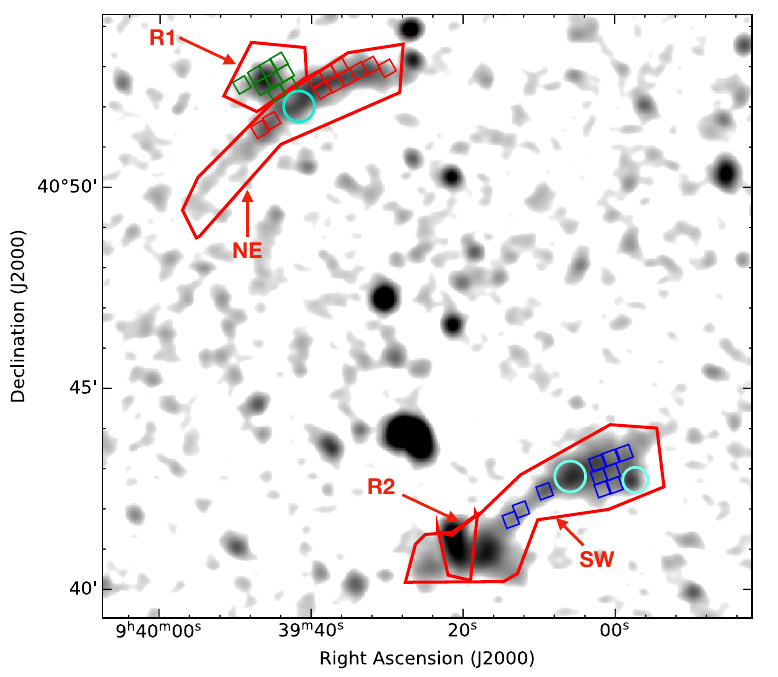}
    \caption{\textit{Left}: Integrated spectra of diffuse radio sources in the PSZ2\,G181.06+48.47 cluster from 144\,MHz to 1.5\,GHz, well-described by single power-laws (dashed lines). \textit{Right}: LOFAR 144\,MHz image overlaid with regions used for extracting flux densities of NE, SW, R1, and R2. Cyan circles mark masked point sources. Boxes are regions (where emission is detected at all the observed frequencies) used for radio color-color analysis.}
    \label{fig:spectrum}
\end{figure*}

Polarization intensity maps with overlaid magnetic field vectors focusing on the NE and SW regions are shown in Figure~\ref{fig:vector_maps}, while the resulting RM map is shown in Figure~\ref{fig:RM_map}.

\section{Results}
\label{sec:results}

We investigate a number of diffuse sources in the cluster, whose properties are summarized in Table~\ref{tab:flux}.

\subsection{Radio morphologies}

Two large diffuse sources located on the northeast and southwest periphery of the cluster (NE and SW) are detected across 1.2\,dex in frequency, from 120\,MHz to 2\,GHz (Figures\,\ref{fig:high_res} and \ref{fig:low_res}). Measured from center to center, their projected relative distance is $\sim2.6$\,Mpc. 

With a largest linear size (LLS) of $\sim$1.2\,Mpc at 144\,MHz and a very thin ($\sim70$\,kpc) morphology towards its northeastern end, the NE source appears to consist of two filaments, in particular at 1.5\,GHz (Figure~\ref{fig:high_res}). The source is located about 1.35\,Mpc from the system center. The sensitive high-resolution images allow us to identify at least seven compact sources with optical counterparts embedded in NE, with the one source with a spectroscopic redshift ($z_\mathrm{spec}=0.499584$) confirmed as not a cluster member. We present a point-source-subtracted high resolution Band\,3 image of the NE source in Figure~\ref{fig:NE_R1_overlay}. The diffuse (LLS\,=\,370\,kpc at 144\,MHz) source R1 is connected to NE via a faint, 125\,kpc-wide bridge of emission.

Source SW, located $\sim1.18$\,Mpc away from the cluster center, has a non-uniform brightness across its 1.3\,Mpc extension (at 144\,MHz), which peaks towards the west with a downstream width of about 420\,kpc at 144\,MHz and 280\,kpc at 1.5\,GHz (Figure\,\ref{fig:low_res}). At the center, the source is much thinner, with a width of $\sim90$\,kpc. We identify five embedded compact sources, none of which has spectroscopic redshift measurements (see Figure~\ref{fig:SW_R2_overlay}). At SW's eastern end, we detect a blob of bright tailed emission (R2). While R2 appears connected to the eastern end of SW, R2 extends further to the east. The uGMRT Band\,4 high-resolution image separates the source into a compact core and a faint, extended tail (Figure\,\ref{fig:high_res}). The core is only marginally detected at 1.5\,GHz suggesting a steep spectrum. A spectroscopically-confirmed cluster member ($z_\mathrm{spec}=0.235086$) coincides with the compact core emission at the north of R2, as shown in Figure~\ref{fig:SW_R2_overlay} (left panel).

\begin{figure*}[!thbp]
    \centering
    \includegraphics[width=0.47\textwidth]{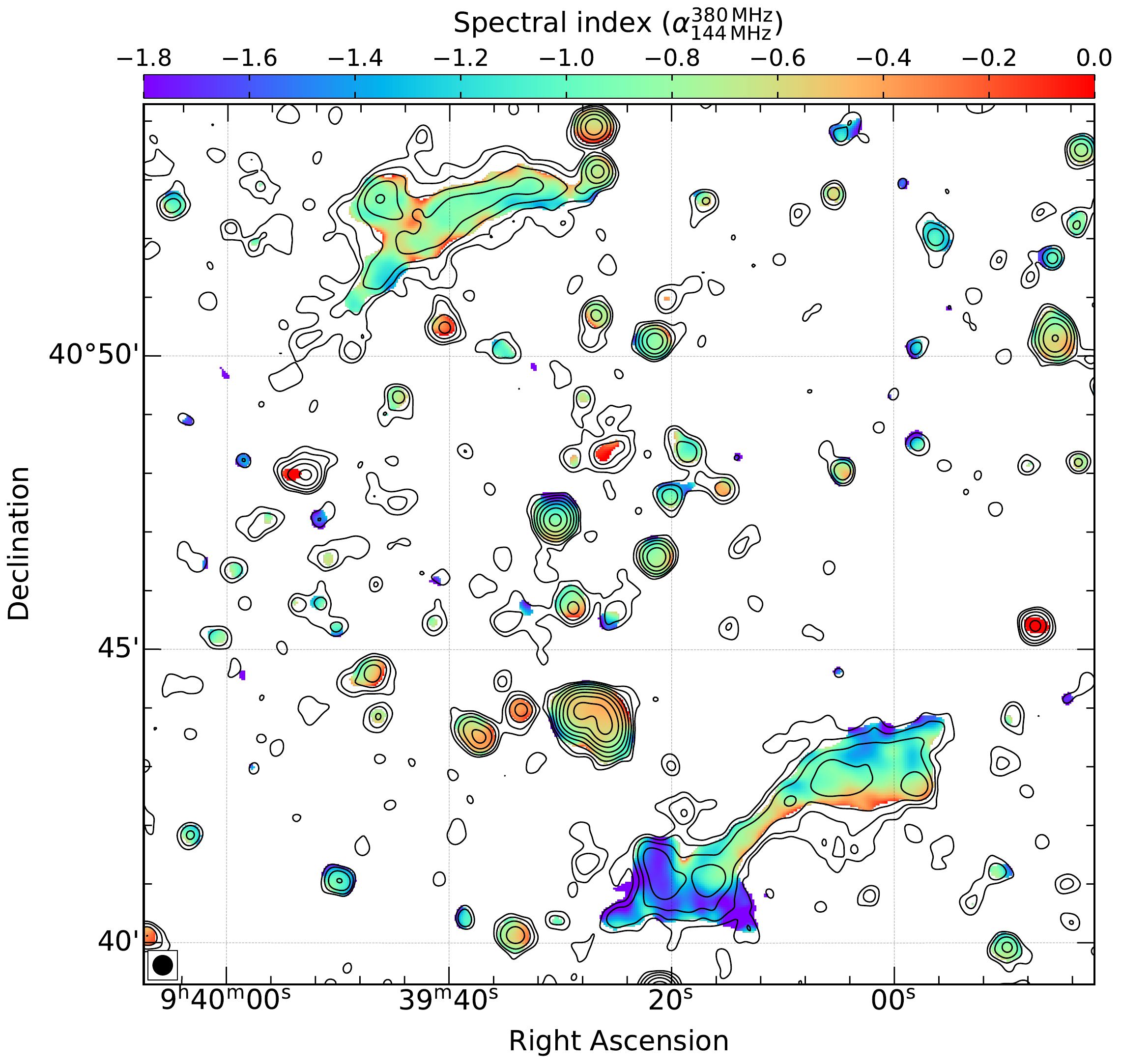}
    \includegraphics[width=0.47\textwidth]{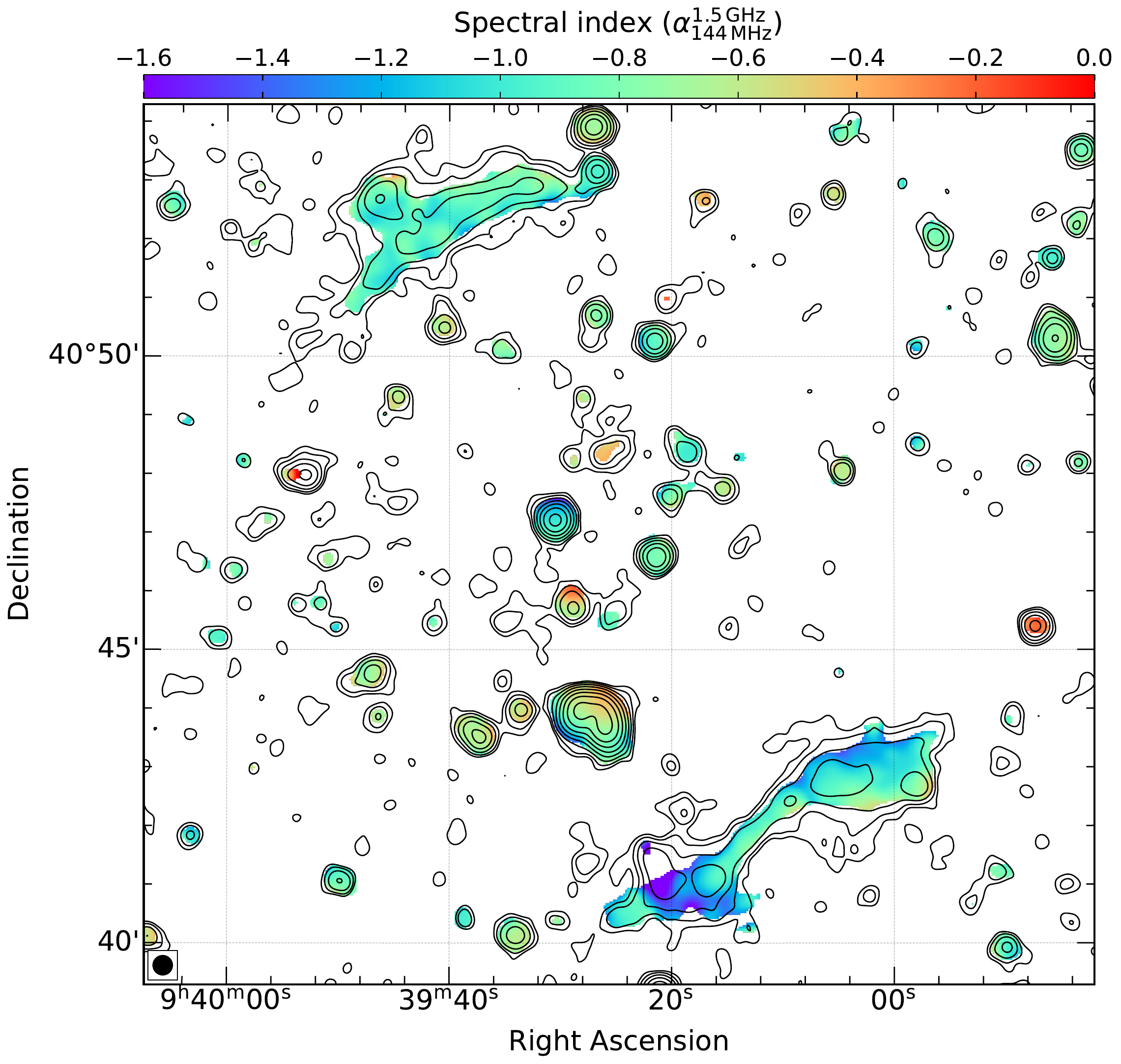}
    \caption{Low and high frequency spectral index maps at 20{\arcsec}-resolution: 144\,MHz and 380\,MHz (left) and 144\,MHz and 1.5\,GHz (right). Stokes I contours from the uGMRT Band\,3 uv-tapered images are plotted at [1, 2, 4, 8...] $\times 3\,\sigma_\textrm{rms}$ where $\sigma_\textrm{rms}=23$\, $\mu$Jy\,beam$^{-1}$.}
    \label{fig:index_maps}  
\end{figure*}

\begin{figure}[!thbp]
    \centering
    \includegraphics[width=0.48\textwidth]{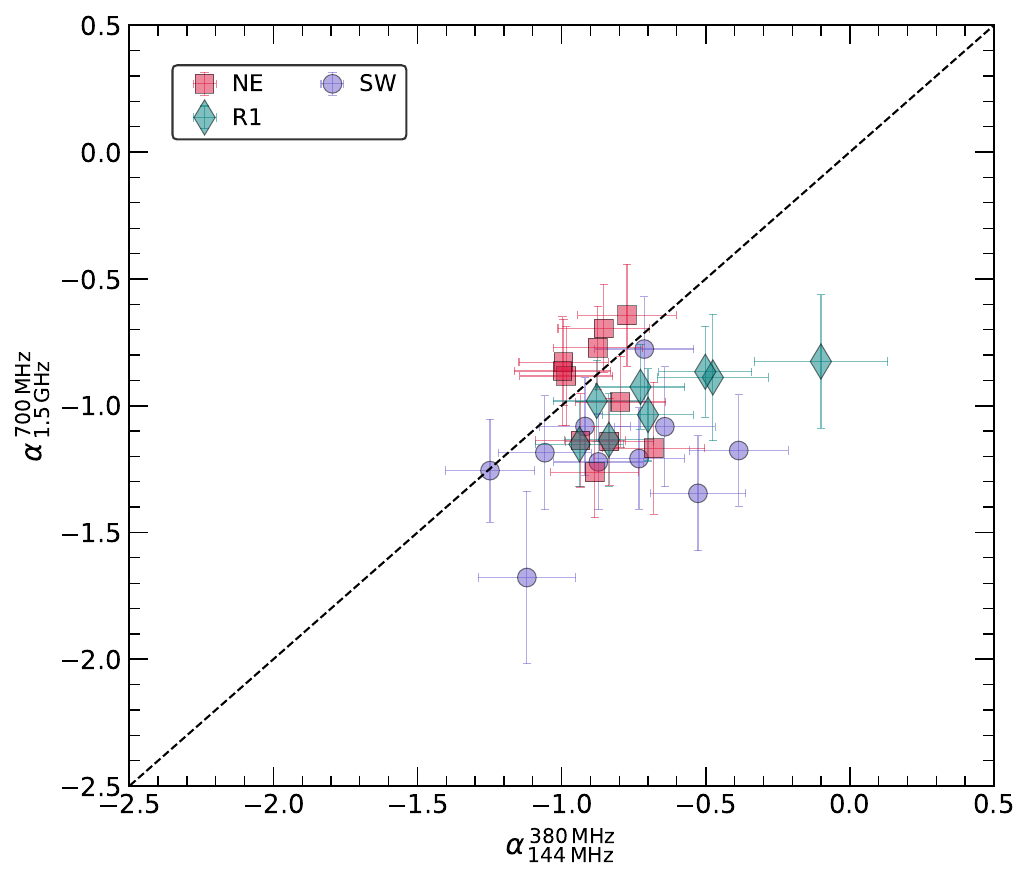}
        \vspace{-0.7cm}
    \caption{Radio color-color plot of the NE relic, SW relic and R1, created using the flux densities extracted from 20\arcsec boxes (i.e., the beam size), corresponding to a physical size of about 76\,kpc, as shown in the right panel of Figure~\ref{fig:spectrum}. The dashed line implies where $\alpha^{380\rm\,MHz}_{144\rm\,MHz}$=$\alpha_{700\rm\,MHz}^{1.5\rm \,GHz}$. The three sources display a range of behavior, with only SW showing evidence for spectral aging.}
    \label{fig:cc_plot}  
\end{figure}

In our lowest resolution GMRT Band\,3 image, we observe a hint of low-significance ($\geq2\sigma$ level) diffuse emission in the central region of the cluster (discussed in Section\,\ref{sec:halo}). The source is not detected at a similar level in any of the other low-resolution maps, suggesting a steep spectrum. 

\subsection{Integrated radio spectra and radio powers}
\label{sec:radio_spec}

The integrated spectra of NE, SW, R1, and R2 between 144\,MHz and 1.5\,GHz are shown in Figure~\ref{fig:spectrum} left panel and summarized in Table~\ref{tab:flux}. Note that the spectral indices are measured after excluding the compact sources with optical counterparts. The integrated spectral index of NE and SW are $\alpha=-0.92\pm0.04$ and $\alpha=-1.09\pm0.05$, respectively. The radio emission between NE and SW has a flux density of about 2.1\,mJy at 380\,MHz, excluding discrete sources (marked in Figure\,\ref{fig:spectrum} right panel). It is worth mentioning that the LOFAR image suffers from calibration artifacts in the central region, resulting in negative flux density, thus not allowing us to recover an accurate flux density at 144\,MHz. 

We obtain the monochromatic radio power as:
\begin{equation}
\label{eq:power}
P =4\pi D^{2}_{L}S_{\nu_{i}} (1+z)^{-(1+\alpha)},  
\end{equation}
where $D_{L}$ is the luminosity distance to the source, $\alpha$ is the spectral index used in the $k$-correction, $S_{\nu_{i}}$ is the flux density at 1.5\,GHz and 144\,MHz. The estimated radio powers for the diffuse sources in the cluster are listed in Table~\ref{tab:flux}. All the diffuse sources are adequately fit with a single power-law up to 1.5\,GHz without any obvious break. 

\subsection{Spatially-resolved spectral index maps}

The NE source has no apparent spectral index gradient  either high or low resolution  spectral index maps (see Figure~\ref{fig:NE_R1_overlay} right panel and Figure~\ref{fig:index_maps}). We observe instead varying spectral indices across it, mainly varying between $-0.55$ and $-0.9$. Similarly, no clear spectral index gradient oriented toward the cluster center is seen in the emission of diffuse source R1. 

A noticeable spectral index gradient is observed for the SW source with values varying from $-0.6$ at the southern leading edge to $-1.6$ at the northern, see Figure~\ref{fig:index_maps}. R2 has distinctly steeper spectral indices compared to the bulk of SW, with the steepest spectral index across R2 reaching $-1.8$. The high-resolution spectral index map unveils a steep spectral index in the core of R2 (about $-1.0$), followed by radial steepening in the southwest direction (right panel in Figure~\ref{fig:SW_R2_overlay}).

\subsection{Polarization properties}

\begin{figure*}[!thbp]
    \centering
    \includegraphics[width=0.48\textwidth]{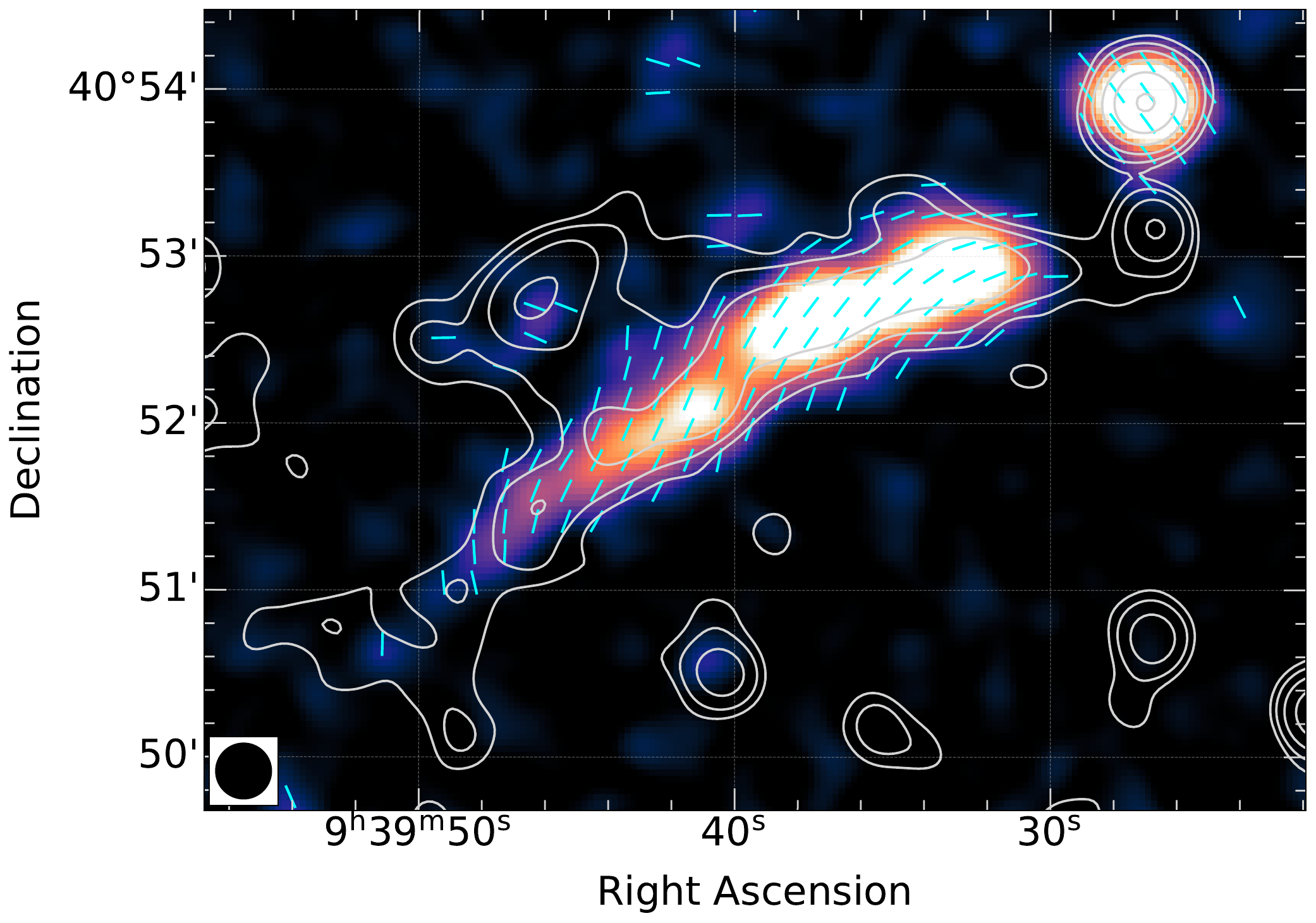}
    \includegraphics[width=0.45\textwidth]{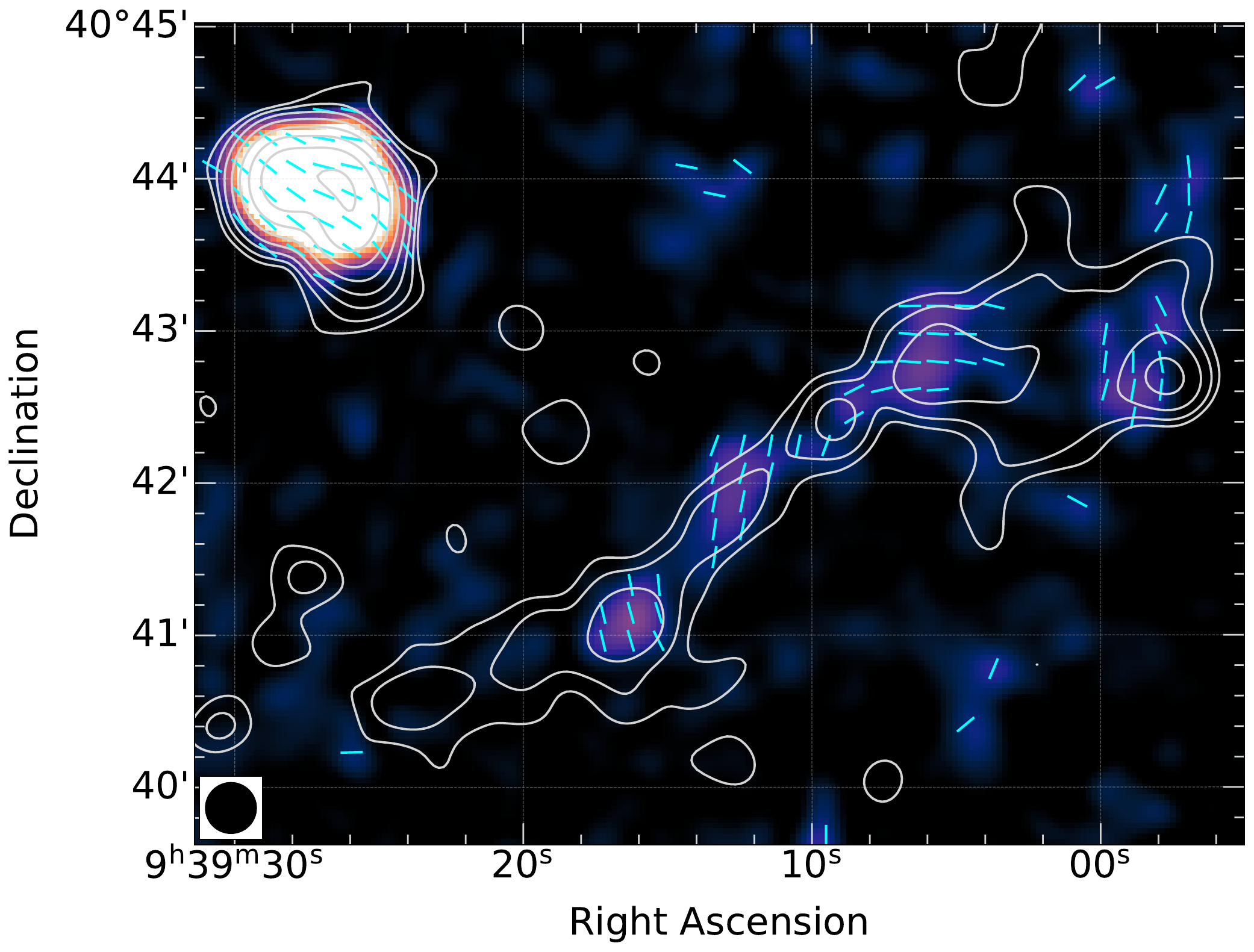}
    \vspace{-0.3cm}
    \caption{Polarization intensity maps of the NE (left) and SW (right) relics overlaid with magnetic field vectors. Their orientation represents the projected magnetic field corrected for contribution from the Galactic foreground. Stokes I contours from the 1.5\,GHz VLA image are drawn at $[1, 2, 4, 8 ...]\times 5\sigma_{\rm rms}$. The two relics are very differently polarized.}
    \label{fig:vector_maps}
\end{figure*}

As depicted in Figure~\ref{fig:vector_maps}, the NE source exhibits polarization throughout its entire extent, with the degree of polarization varying between 5\% to 42\%, with an average of 22\%. The magnetic field vectors, after correcting for the Faraday rotation, are aligned with the long axis of the NE source. Unlike NE, R1 is almost completely depolarized at L-band.

Some parts of the SW source are polarized, with fractional polarization in the 1--20\% range and an average polarization of 14\%, and with magnetic fields aligned with the long axis. R2 is not detected in polarization.

\subsection{RM distribution and magnetic field estimates}

The distribution of the peak of RM, the so-called ``rotation measure" map is shown in Figure~\ref{fig:RM_map}. The RM spatially varies in the NE source mainly from about +10 to $+35\,\rm rad \,m^{-2}$, but there are also a few patches with RMs ranging from about $\rm +300\,rad\,m^{-2}$ and $\rm -140\,rad\,m^{-2}$. The average $\sigma_{RM}=16\,\rm rad \,m^{-2}$ across NE suggests very little Faraday rotating intervening material. For the SW source, we only detect a few patches where the RM is mainly in the range +6 to +15\,$\rm rad \,m^{-2}$. Since the expected Galactic RM at the location of the cluster is $\sim3\,\rm rad\,m^{-2}$ \citep{Hutschenreuter2022}, the detected RMs slightly deviate from the average Galactic foreground.

Under a simple assumption, the dispersion in RM within the telescope beam, $\sigma_{RM}$, can be used to constrain the magnetic field strength \citep{Sokoloff1998}:
 \begin{equation}
  \sigma_{\rm RM} =
  \sqrt{\frac{1}{3}}\cdot 0.81 \cdot  \frac{\langle n_{\rm e}\rangle}{\rm cm^{-3}} \cdot \frac{\,B_{\rm turb}}{\rm \mu G} \cdot \sqrt{\left(\frac{L}{\rm pc} \cdot \frac{t}{\rm pc}\right) \cdot \frac{1}{f}} ,
\label{sigmaRM}
\end{equation}
where $\langle n_{\rm e}\rangle$ represents the average thermal electron density of the ionized gas along the line-of-sight, $B_{\rm turb}$ denotes the magnetic field strength, and $f$ represents the volume filling factor of the Faraday-rotating plasma. $L$ and $t$ refer to the path length through the thermal gas and turbulence scale, respectively.

We use $f = 0.5$ \citep{Murgia2004}, $t=60$\,kpc (RM fluctuations) and $\sigma_{RM}=16 \rm\,rad\,m^{-2}$ for the NE relic. We assume a thermal electron density at $R_{500}$ of $n_{e}=10^{-4} \rm\,cm^{-3}$ (the value cannot be measured directly from the data because of the low S/N), and $L\sim1.5$\,Mpc. By inserting all values in Equation\,\ref{sigmaRM}, we obtain a value of $B=0.8\,\rm \mu G$ for the NE relic. We note that the obtained value depends of the assumed $L$ and the magnetic field value should be seen as a lower limit. The strong depolarization across the SW relic does not allow us to constrain its intrinsic polarization fraction, $\sigma_{RM}$ and thus the magnetic fields. However, some detected polarized patches suggest a similar magnetic field value for the SW relic. 

\section{Discussion}
\label{sec:discussion}

PSZ2\,G181.06+48.47 provides an ideal environment to test particle acceleration in a low-mass cluster and a low plasma density regime. By bringing together different lines of evidence from both the radio data and the complementary X-ray and WL observations, we aim to obtain a comprehensive understanding of the cluster by exploring the origin of the multitude of diffuse radio sources from the perspective of the cluster merger scenario.

\vspace{-1.2em}
\begin{deluxetable*}{ccccccc}[!thbp]
\tablecaption{Diffuse radio source properties. We include the classification, the average and maximum polarization fraction, the radio-derived and X-ray-derived Mach number, and magnetic field constraints.}
\tablehead{
Source & Classif. & Avg. pol. & Max. pol.&  $\mathcal{M}_{\rm R}$ & $\mathcal{M}_{\rm X}$ & $B$ \\
 & & frac. & frac. & & & $\mu$\,G \\
}
\startdata
NE & relic & 22\% & 42\% & $>1.7$$^{\dagger}$ & $<1.43$ &  0.8 \\
SW & relic & 14\% & 20\% & $4.8^{+2.3}_{-0.9}$$^{\ddagger}$ & $<1.57$ & \\
R1 & ? & $-$& $<4$\% & \\
R2 & radio galaxy & unpol. & unpol. & \\
H & halo & -- & -- & \\
\enddata
\tablecomments{$^{\dagger}$Constraint from polarization. $^{\ddagger}$Measured from spectral index.}
\label{tab:classification}   
\end{deluxetable*} 
\vspace{-1.2em}

\subsection{Diffuse radio source classification}

The first step in getting a comprehensive picture of the diffuse emission in the cluster is elucidating the nature and classifying sources NE, SW, R1, and R2.

Based on its location relative to the center of the cluster, its morphology, high degree of polarization (discussed in Section\,\ref{sec:polarization_discussion}) and the absence of clear optical counterparts, we classify NE as a radio relic. The integrated spectrum of the source is in line with other relics, which follow a remarkable power-law distribution over a large frequency range \citep{Loi2020, Rajpurohit2020b, Rajpurohit2022b, Murgia2024}. Our estimated radio power for NE (see Table~\ref{tab:flux}) is also in agreement with the radio power versus mass relations of known relics \citep{DeGasperin2014, Jones2023, Duchesne2024}.

The diffuse emission, steep-spectrum, large-scale emission in the SW source points towards a relic classification. None of the five embedded compact sources (Figure~\ref{fig:SW_R2_overlay}) have spectroscopic redshifts, but there seems to be no direct morphological connection with the relic emission. The SW relic morphology in the high-resolution maps (see Figure~\ref{fig:SW_R2_overlay}) is inconsistent with that of a double lobe radio galaxy: there is no obvious compact core emission bracketed by lobe/plume-like emission, the implied physical size places the source in the rare giant-radio-galaxy category \citep{Dabhade2020}, and no bright, optical host can be identified. The spectral index gradient across the SW \citep[e.g.,][]{vanWeeren2011, Stroe2013, Rajpurohit2018} supports a relic origin and rules out its classification as a double-lobed radio galaxy. The leading southern edge of the relic is the site of CR electron (CRe) acceleration at the shock front, where we expect the flattest spectral index. In the downstream regions, the spectral index steepens as the shock front propagates outwards, which is believed to be due to the aging of electrons because of inverse Compton and synchrotron energy losses \citep{vanWeeren2010, Stroe2013} or variation of the Mach number across the shock surface \citep{Skillman2013}. Further, SW's radio power falls confidently within the expected relic scaling relations \citep{DeGasperin2014, Jones2023, PSZ_Xray}. For the  PSZ2\,G181.06+48.47 relics and other known double relics radio power versus mass relations, we refer to \cite{PSZ_Xray}.

Despite their clear connection, the observations strongly suggest a different nature for R2 compared to SW. R2's morphology resembles a tailed radio galaxy, in particular at 144\,MHz and 380\,MHz. Moreover, the spectral index of $-1.0$ at R2's core is much steeper than values typically detected in the core of radio galaxies \citep[mostly in the range of $-0.5$ to $-0.7$;][]{Stroe2013, Botteon2020}, which seems consistent with an old tailed radio galaxy. 

On the basis of radio morphology, extent, and flat spectral index, R1 could be a radio galaxy, however, we did not find an optical counterpart. The flat spectral index implies that R1 is not a radio phoenix produced by adiabatic compression of fossil radio plasma by the passage of an ICM shock front nor an AGN relic, originating from a fossil population caused by previous AGN activity, both of which are characterized by steep spectral indices \citep{Ensslin2001, Ensslin2002, Mandal2020}. 

\subsection{Mach numbers for the radio relics}

For a (quasi)-stationary state shock (where the cooling time of electrons is much shorter than the timescale on which the shock strength changes), assuming a standard power-law energy distribution of relativistic electrons just after acceleration, the integrated spectral index ($\alpha_{\rm int}$) is steeper than the injection index ($\alpha_{\rm inj}$):
\begin{equation}
\alpha_{\rm int}=\alpha_{\rm inj}-0.5.
\label{eq:stationary}
\end{equation}

In the DSA model in the test-particle regime, the radio-integrated index is related to the Mach number ($\mathcal{M}$) of the shock as \citep{Blandford1987}: 
\begin{equation}
\mathcal{M}=\sqrt{\frac{\alpha_{\rm int}-1}{\alpha_{\rm int}+1}}.
\label{eq:int_mach}
\end{equation} 
If generated according to the quasi-stationary shock scenario \citep{Kardashev1962}, 
radio relics are expected to show a power law with spectral index $\leq -1$ (see Equation~\ref{eq:stationary}).

The SW relic in PSZ2\,G181.06+48.47 is consistent with following the stationary state shock condition. Using Equation\,\ref{eq:int_mach} and an integrated spectral index of $-1.09\pm0.05$, we obtain a Mach number of $\mathcal{M}=4.8^{+2.3}_{-0.9}$. For the majority of radio relics, X-ray inferred Mach numbers are found to be below 2.5 \citep{Botteon2018, Botteon2020, Wittor2021}, whereas radio-derived Mach numbers are above 2.5. Recent simulations indeed suggest that shock fronts have a distribution of Mach numbers and that the Mach numbers obtained from radio-integrated spectra are biased toward the high Mach number tail of the distribution \citep{Paola2020a, Wittor2021}.

The stationary state shock condition does not hold for the NE relic because its overall spectrum is quite flat ($\alpha_{\rm int}=-0.92\pm0.04$). This spectrum, considering the stationary shock condition, would imply an injection spectral index $-0.42$, 0.08 flatter than the flattest allowed by the DSA model. Even though Mach number predictions are not possible at these flat spectra assuming, only a relatively strong shock will produce such a flat spectrum. Simulations consistently show that the stationary shock conditions do not hold for spherically expanding shocks \citep{Kang2016a} or within a turbulent medium \citep{Paola2020a}.

\begin{figure*}[!thbp]
    \centering
    \includegraphics[width=0.48\textwidth]{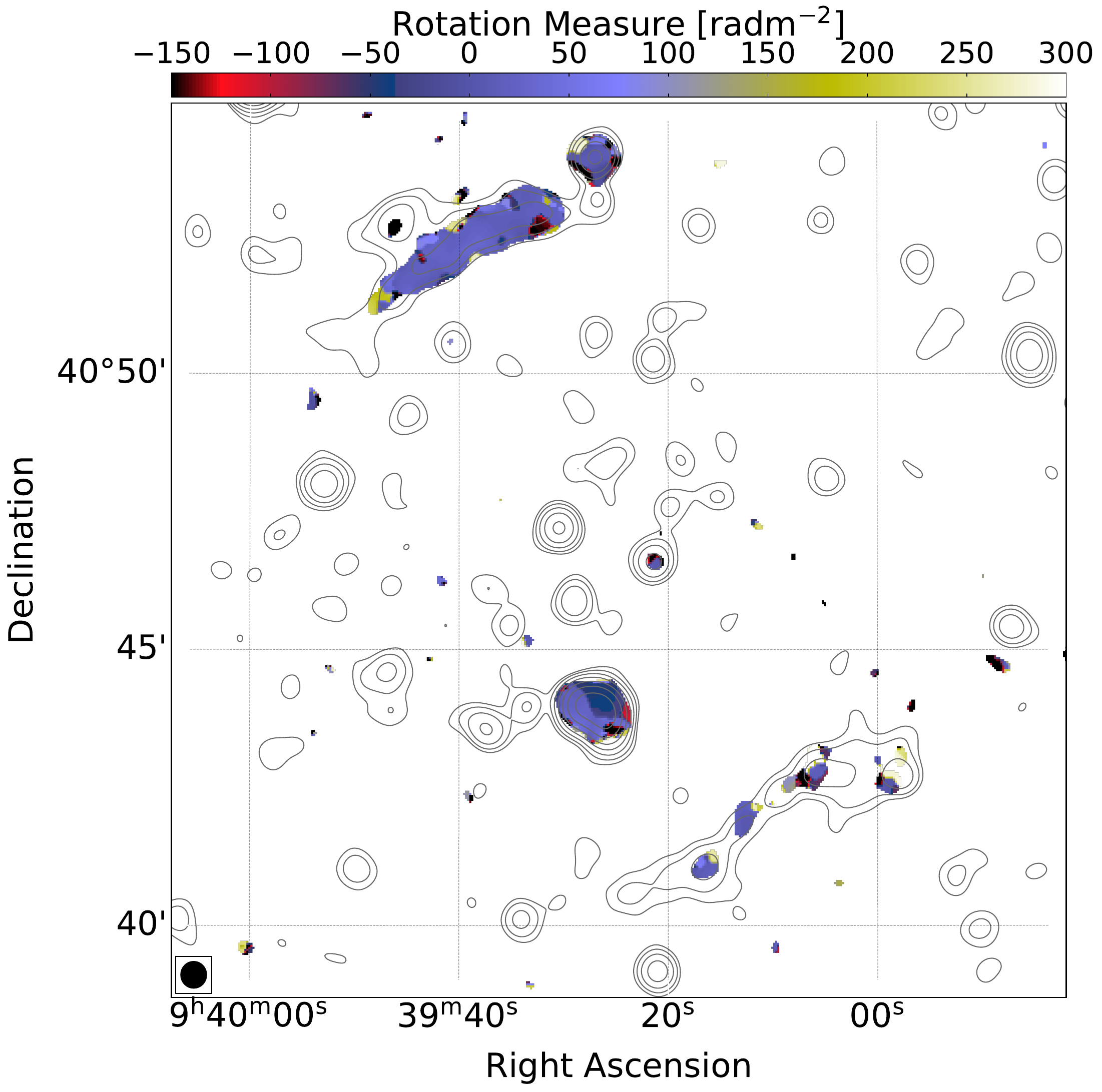}
    \includegraphics[width=0.48\textwidth]{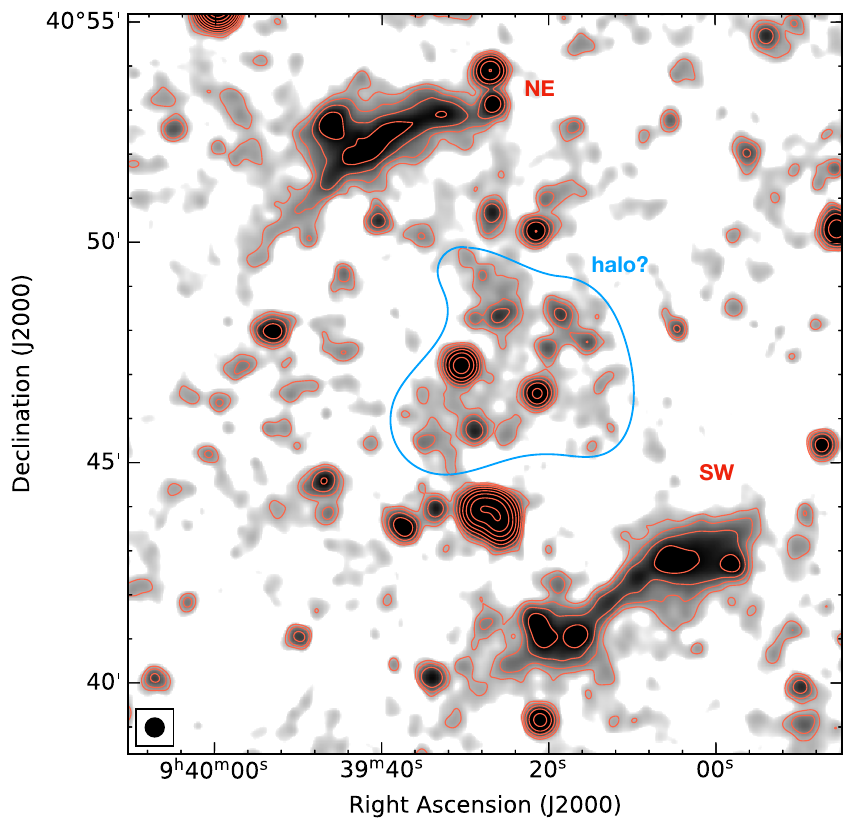}
    \caption{\textit{Left:} Rotation Measure map measured over 1.0--2.0\,GHz using the RM-synthesis technique. The beam size is 25{\arcsec}. The map is corrected for the Ricean bias. \textit{Right:} uGMRT Band\,3 total intensity image of the cluster at 25\arcsec resolution. Red contours are drawn at [1, 2, 4, 8, ...] $\times 3\,\sigma_\textrm{rms}$, where $\sigma_\textrm{rms}=40$\, $\mu$Jy\,beam$^{-1}$ and from uGMRT Band\,3 image (25\arcsec resolution). The candidate radio halo is marked in blue.} 
    \label{fig:RM_map}
\end{figure*}

An alternative method to determine the Mach number is measuring the injection index directly at the relic's leading edge from high-resolution spectral index maps. It is worth mentioning that the combination of projection effects, cooling distance, and smoothing in the radio images impacts the estimation of the injection index. Moreover, high resolution studies of bright relics reveal that the injection index varies across the leading edge of the relic, which in turn implies the Mach number varies across the shock front as found in simulations \citep{Rajpurohit2018, Hoang2018, Wittor2021}. We emphasize that the integrated Mach number is based on the emission-weighted spectral index distribution of the entire relic, i.e., higher Mach numbers have more weight. Mach numbers based on injection spectral index show significant variations along the relic. Therefore, the flattest injection Mach number is expected to be comparable to the integrated Mach number. 

The flattest spectral index at the leading edge of the SW relic is about $-0.6$, which implies a shock of Mach number $\mathcal{M}\sim 4.5$ (using Equations\,\ref{eq:stationary} and \ref{eq:int_mach}). This agrees well with the shock strength obtained from the integrated spectral index. The narrow width of the NE relic makes it difficult to measure the injection index across its leading edge. 

\subsection{Resolved spectral properties}

A single spectral shape (or trajectory) in the color-color plot (Figure~\ref{fig:cc_plot}) throughout the emitting region suggests that every line of sight probes a single electron population and that conditions along each line of sight are relatively homogeneous \citep{KatzStone1997, Rajpurohit2021a}. If data points show positive curvature, that suggest either emitting regions are overlapping, or that they have inverted spectra. Unlike some well-studied radio relics, for example CIZA\,J2242.8+5301 \citep{Stroe2013, Gennaro2018}, RX\,J0603.3+4214 \citep{vanWeeren2012a, Rajpurohit2020a}, Abell\,2744 \citep{Rajpurohit2021c}, we do not find a single spectral shape or trajectory in the color-color plane for the NE and SW relics. 

The majority of the data points from the NE relic are clustered around the power-law line, implying negligible aging, in contrast to simple expectations of strong aging behind shock fronts. The R1 data points don't follow the same trend as the NE relic. This finding hints at the presence of distinct electron populations in these two regions, indicating that R1 and the NE relic may have different origins.

For the SW relic, we see a large scatter in the color-color plot. The color-color plots of simulated relics suggest that the locus of points changes significantly with the relic viewing angle\footnote{The viewing angle is defined such that a source parallel to the plane of the sky and viewed face-on has a viewing angle of $0\degree$, meaning the plane of the object is parallel to the plane of the sky. A source perpendicular to the plane of the sky and viewed edge-on has an viewing angle of $90\degree$} \citep{Rajpurohit2021a, Wittor2019}. In a face-on view, the relativistic electron distribution shows a wide scatter in the color-color plane across the relic emitting region. However, in an edge-on view, the distribution follows a single spectral shape, indicating relatively uniform conditions along each line of sight. Therefore, the viewing angle of the SW and NE relics, on the basis of radio color-color plot trajectory \citep{Rajpurohit2021a}, is very likely inclined along the line of sight ($\sim45\degree$).

\subsection{Polarization insights}
\label{sec:polarization_discussion}

While both relics show well-aligned magnetic field vectors, the NE relic is significantly more polarized than the SW relic. 

The high polarization fraction measured within the NE relic is in line with several radio relics that have been observed to exhibit a local polarization fraction of up to 65\% \citep[e.g.,][]{Bonafede2014, Owen2014, Loi2020, deGasperin2022, Rajpurohit2020b, Rajpurohit2022a}. The fractional polarization of radio relics is predicted to rise with increasing shock Mach numbers. According to the \cite{Ensslin1998} model, for a relic seen perfectly edge-on and assuming no further depolarization, the average fractional polarization is anticipated to increase from 40\% to 60\% when the Mach number increase from $\mathcal{M}=1.5$ to $\mathcal{M}=3$. Using a numerical model, \cite{Hoeft2022} and \citet{Kierdorf2017} found similar trends. For the NE relic, our observation shows that, at 1.5\,GHz, the polarization fraction reaches values as high as 42\% at 25\arcsec resolution, implies that the minimum Mach number in that region is 1.7 according to \citet{Kierdorf2017, Hoeft2022}.

The RM distribution across the NE relic implies very little Faraday Rotating intervening material and the intrinsic polarization angle is well aligned with the orientation of radio emission. A possible explanation for the depolarization in the SW relic is the presence of a Faraday rotating medium (i.e., the ICM). Interestingly, \xmm images reveal faint diffuse X-ray emission at the location of R1 which is strongly depolarized between 1-2\,GHz, suggesting that it may be situated behind or deep inside the ICM and thus is a separate feature and not physically connected to the NE relic. However, high frequency polarization observations are required to confirm this scenario.

The average polarization fraction of radio relics is related to the viewing angle \citep{Ensslin1998} as:
\begin{equation}
    <P_{\rm weak}>=\frac{(s+1)}{(s+\frac{7}{3})} \frac{\sin^2\theta}{\frac{2C^2}{C^2-1}-\sin^2\theta},
    \label{viewing_angle}
\end{equation}
where $<P_{\rm weak}>$ is the mean observed polarization fraction, $s$ is the slope of the electron energy distribution ($\alpha=\frac{1-s}{2}$), $\theta$ is the viewing angle (defined as $90\degree$ for a relic seen edge-on, i.e. perpendicular to the plane of the sky), and $C$ is the compression factor, which is related to the spectral index by:
\begin{equation}
C=\frac{\alpha-2}{\alpha+\frac{1}{2}}.
\end{equation}

Using Equation~\ref{viewing_angle}, we place lower limits on the viewing angle as $\theta\geq 42\degree$ and $\theta\geq 36\degree$ for the NE and SW relics, respectively. The actual values are likely to be higher due to beam depolarization and wavelength dependent depolarization (i.e., intrinsic and external) in the 1--2\,GHz frequency range. While the polarization constraints on the viewing angle exclude extreme scenarios in which the relics are viewed face-on, the merger might not be happening in the plane of the sky, but rather at an inclination $\sim45\degree$ with respect to the plane of the sky.

\subsection{A candidate halo}
\label{sec:halo}
We investigate the potential presence of a radio halo in PSZ2\,G181.06+48.47. As shown in the right panel of Figure,\ref{fig:RM_map}, we find a low-significance detection of diffuse emission towards the cluster center in the low-resolution uGMRT 380\,MHz image. It has a LLS of about 1.1\,Mpc ($2\sigma$). The total flux density of the central diffuse emission is $S_{\rm 380\,MHz}=2.1\pm0.3\rm \,mJy$ ($2\sigma$) and $S_{\rm 700\,MHz}<0.6$\,mJy ($2\sigma$ upper limit). We can therefore, set an upper limit on the spectral index of $\alpha_{380\,\rm MHz}^{700\,\rm MHz} \leq-2.0$. By extrapolating from the radio spectrum, we expect a flux density of about 15\,mJy at 144\,MHz, corresponding to a radio power of $P_{144\,\rm MHz}\leq3\times10^{24}\rm W\,Hz^{-1}$. However, the halo emission is not detected in the LOFAR data. A possible reason for why the expected 144\,MHz flux density is not recovered is due to calibration issue as this region shows negative flux density when images at low resolution. Another possible explanation is that the spectrum of the halo may be curved, i.e., it is flatter below 400\,MHz, as is seen for the radio halo in the Coma cluster \citep{Thierbach2003}, Abell\,S1063 \citep{Xie2020}, and MACS\, J0717.5+3745 \citep{Rajpurohit2022a}. 

Only a handful of radio halos have been found in low-mass clusters ($M_{500}\leq 5\times10^{14}M_{\odot}$), for example, Abell\,1430 \citep{Hoeft2020}, PSZ2\,G145.92-12.53 \citep{Botteon2021}, Abell\,1451 \citep{Cuciti2018}, Abell\,990 \citep{Hoang2021}, and Zwcl\,0634.1+4750 \citep{Cuciti2018}. Detecting such halos is challenging due to the low energy budgets available to inject large-scale cluster turbulence into the ICM. The radio power of the candidate halo in PSZ2\,G181.06+48.47 is in line with the 144\,MHz radio power versus mass relation of known halos \citep{vanWeeren2020, Cuciti2023, Duchesne2024}.

According to the turbulent re-acceleration models, radio halos with ultra-steep spectra are produced by less energetic mergers and in low-mass systems as the cluster mass impacts the energy budget for particle acceleration \citep{Brunetti2014, Cassano2005}. Therefore, the presence of an ultra steep spectrum radio halo in a low mass cluster PSZ2\,G181.06+48.47 is in line with turbulent re-acceleration models. \cite{Donnert2013} investigated the evolution of radio halos during cluster mergers and found that radio halos are brighter and have flat spectral indices ($\alpha\geq-1.3$) in early merger stages, transitioning to fainter halos with steeper spectral indices ($\alpha\leq-1.5$) in later merger stages (after a few Gyrs). The properties of the candidate radio halo PSZ2\,G181.06+48.47 are consistent with it being in an inefficient acceleration stage.

\begin{figure*}[!thbp]
    \centering
    \includegraphics[width=0.45\linewidth]{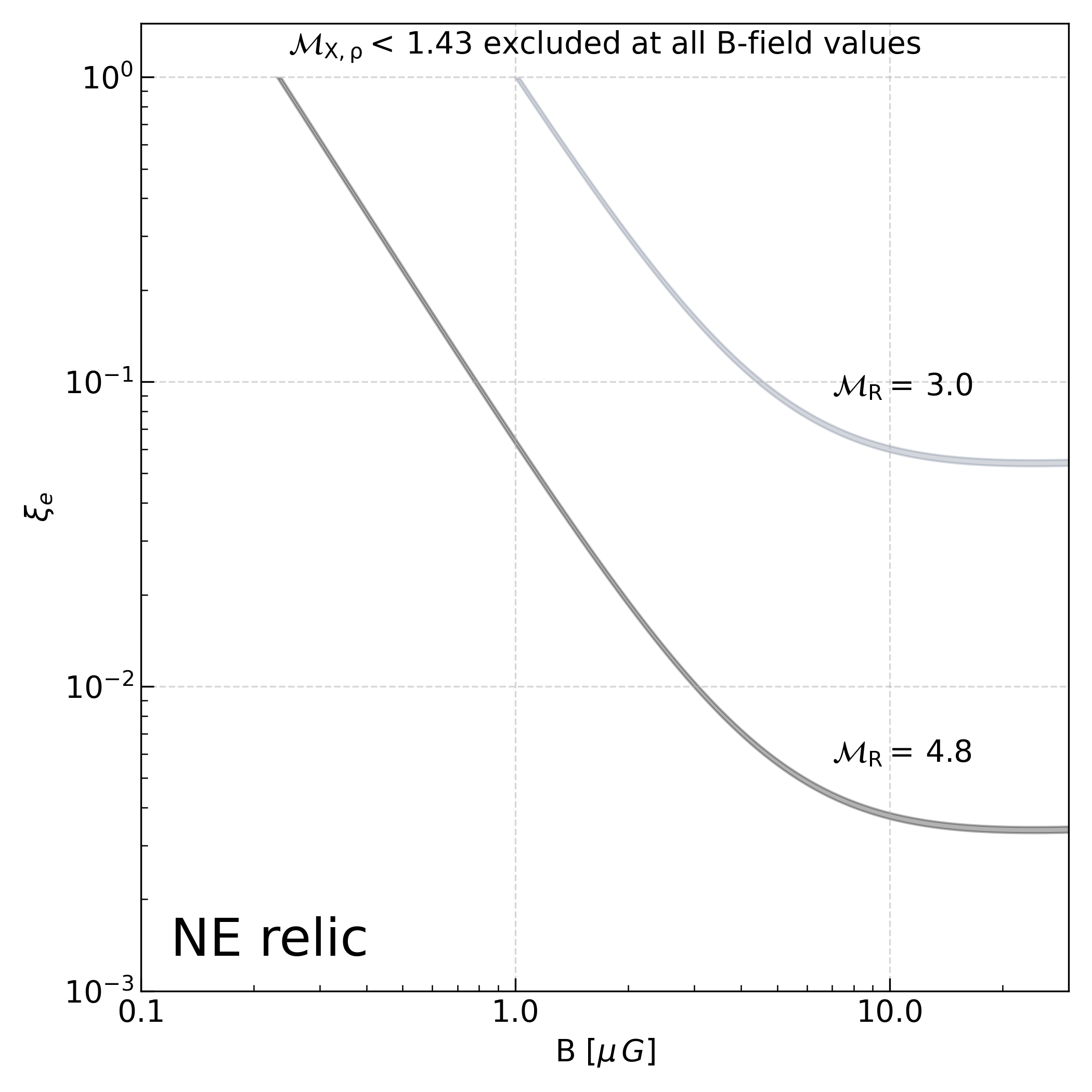}
    \includegraphics[width=0.45\linewidth]{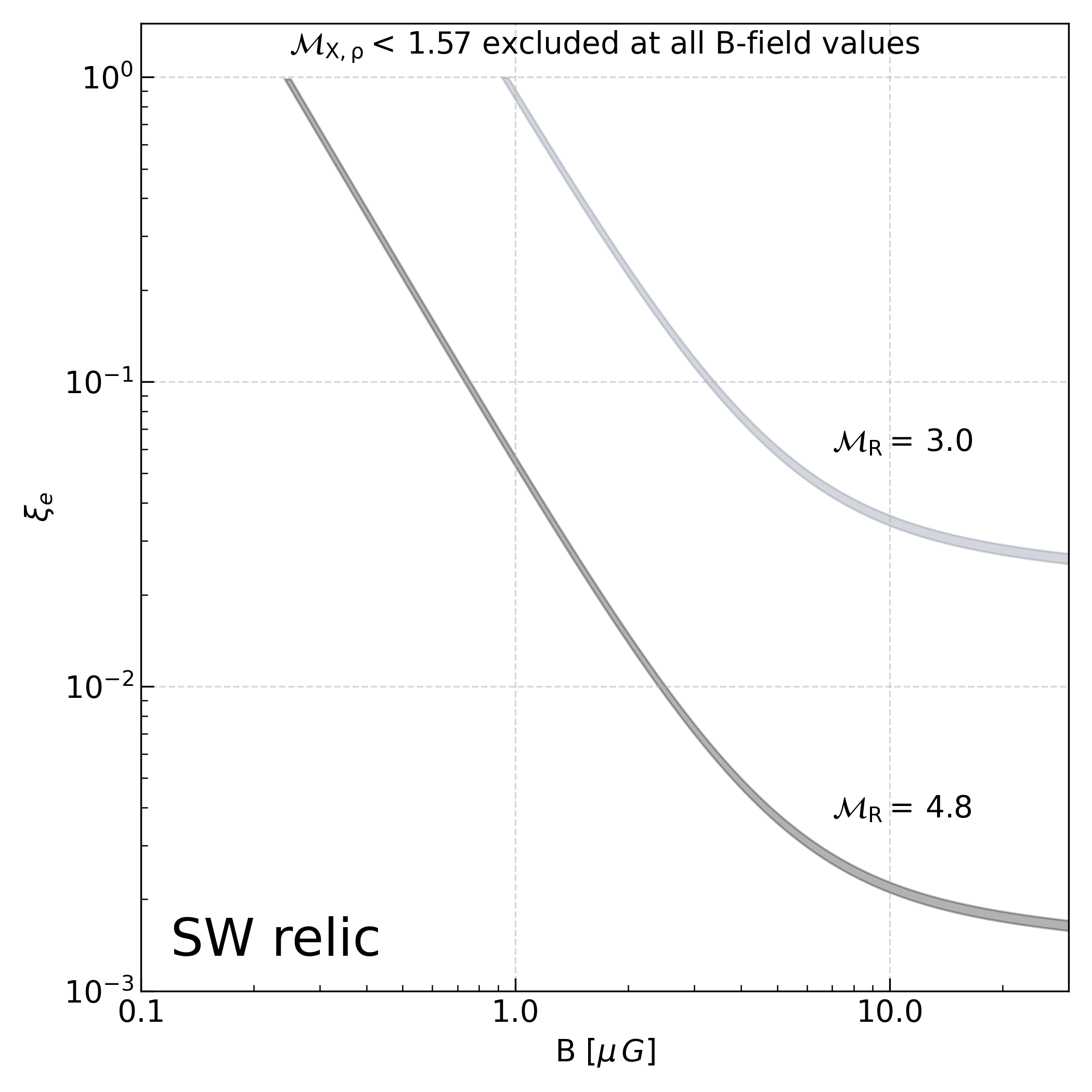}
    \caption{Electron acceleration efficiency as a function of magnetic field in the downstream region. Radio and X-ray Mach numbers are considered. DSA from the thermal pool is compatible with favorable scenarios ($B\gtrsim1$\,$\mu$G and $\mathcal{M}\gtrsim4.8$), but re-acceleration needs to otherwise be invoked to explain the bright and distant relics.}
    \label{fig:efficiency}
\end{figure*}

\subsection{Acceleration efficiency}

We investigate the acceleration efficiencies required to reproduce the observed relic luminosities in a DSA scenario with acceleration by a stationary shock from the thermal pool. We follow the formalism described in \citet{Hoeft2007} to predict the radio relic power from the properties of the downstream plasma. The fraction of the kinetic energy channeled into accelerated thermal pool electrons $\xi_e$ (acceleration efficiency) is computed as:
\begin{equation}
\begin{split}
\xi_e = & 
\frac{1}{C}\cdot \left(\frac{P}{\,{\rm W\,Hz^{-1}}}\right) \cdot \left(\frac{A}{\rm Mpc^3}\right)^{-1} \cdot \left(\frac{n_{\rm e, d}}{10^{-4} {\rm cm^{-3}}}\right)^{-1} \\ 
& \cdot \left(\frac{T_{\rm d}}{7\,{\rm keV}}\right)^{-\frac{3}{2}} \cdot  \left(\frac{\nu}{1.4\,{\rm GHz}}\right)^{-\alpha} \cdot \left(\frac{B}{\rm \mu G}\right)^{1+\alpha} \\
& \cdot \left(\frac{B^2_{\rm CMB}}{B^2}+1\right) \cdot \Psi(\mathcal{M}, T_{\rm d})
\end{split}
\end{equation}
where, $C$ is a constant equal to $1.28\times10^{27}$ \citep{Rajpurohit2021c}, $P$ is the radio power calculated from equation~\ref{eq:power}, $A$ is the relic area calculated as  $\pi/4 \cdot {\rm LLS}^2$  \citep{Hoeft2007, Rajpurohit2021c}, $n_{\rm e, d}$ and $T_{\rm d}$ are the downstream electron density and temperature, respectively, $\nu$ is the observing frequency, $\alpha$ is the integrated radio spectrum, $B$ is the magnetic field, and $B_{\rm CMB}$ is the B field strength equivalent to the Cosmic Microwave Background energy density at the redshift of the cluster ($B_{\rm CMB}=3.24\cdot(1+z)^2$\,$\mu$G). $\Psi(\mathcal{M}, T_{\rm d})$ describes the strong dependency on the Mach number $\mathcal{M}$ and a mild dependency on the downstream temperature $T_{\rm d}$.

We explored the dependency of the acceleration efficiency $\xi_e$ on the magnetic field for a range of B-field values. The radio power $P$ at $\nu=1.5$\,GHz, relic area $A$ and integrated spectral index $\alpha$ are calculated directly from the radio data. We assume $n_{\rm e, d} = 10^{-4}$\,cm$^{-3}$. We use the X-ray measurement of the downstream temperatures from \citet{PSZ_Xray}. Based on these $T_{\rm d}$ values, $\Psi(\mathcal{M}, T_{\rm d})$ was then derived from \citet{Hoeft2007} for a range of Mach numbers, including the radio-derived Mach number for the SW relic ($\mathcal{M}_{\rm R}=4.8$), a typical Mach number for relics ($\mathcal{M}_{\rm R}=3$), and the $5\sigma$ X-ray Mach number upper limits derived from the density jumps in the \xmm data ($\mathcal{M}_{\rm X,\rho}=1.43$ and $\mathcal{M}_{\rm X,\rho}=1.57$ for the NE and SW relics, respectively) \citep{PSZ_Xray}. We show the electron acceleration efficiency required to reproduce the observed radio power of the two relics in Figure~\ref{fig:efficiency}.

Simulations predict cosmic rays acceleration efficiencies from the thermal pool depend strongly on Mach number, with $\lesssim10^{-4}$ for $\mathcal{M}<3$ and reaching $10^{-4}-10^{-3}$ for $\mathcal{M}>3$ \citep{Kang2007, Kang2013}. Under the assumption that CRe are injected from the same thermal population as the CRp, realistic $\xi_e$ values are expected to be $<10^{-2}$, with optimistic values still under 0.1 \citep[e.g.][and references therein]{Botteon2020}. Unrealistically large CRe efficiencies are thus required to reproduce the large observed luminosities of known radio relics given their association with weak shocks ($\mathcal{M}<5$, with most cases $\mathcal{M}<3$) \citep[e.g.][]{Botteon2020}. 

For PSZ2\,G181.06+48.47, the scenarios based on the density-derived Mach number are excluded under all B-field assumptions, as required efficiencies are above 100\%. With compression and amplification, magnetic fields in radio relics are expected to be at the few $\mu$G levels \citep[$1-6$\,$\mu$G, e.g.][]{Finoguenov2010, vanWeeren2010, Rajpurohit2018, Stuardi2021}. Given the $\sim0.6$\,$\mu$G B-field derived from the polarization and assuming amplification factors of a few \citep[e.g.,][]{Donnert2016, Iapichino2012}, the magnetic field in the relics is expected to not exceed a few $\mu$G. Under the optimistic assumption that the B-field is at the few $\mu$G level, neither of the two relics can be reproduced with realistic efficiencies of $\xi_e\sim10^{-2}$. If the magnetic field or Mach number is smaller (e.g., $B<1\,\mu$G and $\mathcal{M}<4.8$), the required efficiency is higher ($\xi_e\gtrsim1$). Further, simulations suggest that Mach numbers measured from the radio observations (as per Equation~\ref{eq:int_mach}) trace the tail of the high Mach number distribution in relics and only a small fraction of the relic area is expected to have such high Mach numbers \citep[e.g.][]{Botteon2020, Wittor2021}, resulting in even more stringent efficiency requirements for the NE and SW relics in PSZ2\,G181.06+48.47 (e.g. lowering the covering fraction from 100\% to 50\% doubles the $\xi_e$ requirement). A patchy surface brightness distribution, like the one we might be observing in NE, could be a sign of an especially high coverage fraction with small Mach numbers \citep{Paola2024}. Therefore, even under the most optimistic assumptions, the relic luminosities can only be reproduced if $\xi_e\sim0.1$ efficiencies are invoked. 

Despite being located above the luminosity-cluster mass relation \citep{PSZ_Xray}, the NE and SW relics are at least $\sim5-50$ times less luminous than bright relics, such as those in CIZA\,J2242.8+5301 \citep{Stroe2013, Gennaro2018}, RX\,J0603.3+4214 \citep{vanWeeren2012a, Rajpurohit2020a}, Abell\,2256 \citep{Rajpurohit2022b}, or Abell\,3667 \citep{deGasperin2022}. However, the combination of the large cluster-centric distances and the low cluster mass results in low ($1-2$\,keV) downstream temperatures, which means that the required efficiencies for NE and SW are still high if particles are accelerated from the thermal pool and if the Mach number of the shock is below 4. In that case, both relics in PSZ2\,G181.06+48.47 are difficult to explain in a simple DSA scenario from the thermal pool. The low acceleration efficiency can be mitigated if some sort of pre-acceleration of CRe is taken into account, which can result in a significant boost of the radio luminosity. That can include, for example, shock re-acceleration of a pre-existing fossil population of seed electrons from AGN activity or re-acceleration by multiple shocks \citep[e.g.][]{Stroe2014, vanWeeren2017a, Kang2021, Inchingolo2022, Smolinski2023}.

\subsection{Origin of the diffuse emission}

The alignment between the elongated X-ray emission \citep{PSZ_Xray}, the WL mass distribution \citep{PSZ_WL} and the two prominent relics to the north and south of the cluster implies a post-core binary merger in the general N-NE to S-SW direction. The large 2.6\,Mpc projected distance between the NE and SW relics prefers a late-stage merger, maybe seen post-apocenter, as the subclusters are moving back towards each other \citep{Hoeft2004, vanWeeren2011, Ng2015, Lee2024}. However, a more detailed analysis of all the observations available reveals a more complicated merger scenario is at play in PSZ2\,G181.06+48.47. The elongated shape with a high aspect ratio between the long and short sides of the relics would naively suggest that they are seen close to edge-on, suggesting the merger is happening in the plane of the sky. However, insights from the polarization and radio color-color plot suggest that the relics are viewed at an inclination $\sim45\degree$ with respect to the plane of the sky. Such a high viewing angle results in a much larger physical separation between the relic of up to $\sim3.5$\,Mpc. To our knowledge, this is the largest relic separation ever found. The possible merger scenarios, including an outgoing merger after the 2nd collision and a returning merger after the first apocenter, are discussed in \cite{PSZ_WL}.

In some simple, binary, head-on collisions, a clear alignment between the orientations of the radio galaxies and the merger axis can be observed \citep[e.g. CIZA\,J2242.8+5301][]{vanWeeren2011, Stroe2013, Jee2015}. During the merger, as galaxies decouple from the ICM, tails of radio galaxies trace their relative motion with respect to the tenuous hot gas. We observe no such clear alignment in PSZ2\,G181.06+48.47. The tails of the two cleanly resolved radio-emitting cluster members, R2 and the prominent wide-angle tailed source almost directly north, are roughly aligned on the N-S direction at an $\sim15\degree$ and $\sim40\degree$, respectively, angle from the merger axis inferred from the ICM, WL and radio relics \citep{PSZ_Xray, PSZ_WL}. However, the morphology and spectral index trends across these two sources indicate that they are both moving roughly northward relative to the ICM in their vicinity. The tailed radio sources seems to be consistent with a late-stage, returning merger scenario, in which the subclusters are falling back towards each other after reaching the maximum separation after the first core passage. 

The NE relic has some features typically associated with relics, including its broad arc-like morphology and high polarization fraction. However, the relic's flat spectral index is inconsistent with the standard DSA from the thermal pool. Moreover, the radio polarization and X-ray analysis suggest a relatively weak shock ($\mathcal{M}\sim1.8$) associated with the NE relic, requiring unrealistic acceleration efficiency if the CRe are accelerated from the thermal pool. A flat spectral index and lack of a clear spectral index gradient behind the shock front across NE, indicate a young population of electrons. While the $\sim45\degree$ viewing angle is difficult to reconcile with the narrow relic width, it can help explain the lack of a clear spectral index trend. The evidence supports the re-acceleration scenario. 

However, none of the AGN embedded in the NE relic appear to be morphologically connected to it, so it is unclear what could be the source of seed electrons. We note that the re-acceleration of ``ghost" plasma (i.e., fossil plasma from an AGN that is not detected with current telescopes due to low energy) is still possible. If the NE relic is formed due to the MSS, a break in the overall spectral is expected at 1.4\,GHz \citep{Inchingolo2022}. Future high-frequency observations would be critical to ascertain the viability of the MSS for the NE relic.

In contrast to the NE relic, the SW relic shows a clear spectral index gradient towards the cluster center and a high polarization fraction, as expected in the DSA model. However, there is a large discrepancy in the radio and X-ray estimated Mach numbers which is reported for several other relics. This mismatch can be explained by the fact that the radio and X-ray observations trace different parts of the Mach number distribution as revealed by simulations.

A re-acceleration scenario is also possible for the SW relic. The radio emission from the SW relic extends to the east of radio galaxy R2. At the location of the relic, the spectral index again flattens, which may suggest that R2 is providing seed  electrons to the relic. A similar change in the spectral index distribution is observed for the relics and nearby radio galaxies in PLCK\,G287.0+32.9 \citep{Bonafede2014}, RXC\,J1314.4-2515 \citep{Stuardi2019}, CIZA\,J2242.8+5301 \citep{Gennaro2018}. However, it is not obvious that the injection of fossil electrons by one radio galaxy can produce a uniform population of electrons forming $\sim1.3$\,Mpc large relic. Moreover, the flatter spectral index observed at the southern tip of the R2 tail could be simply due to projection effects. With the current data, we cannot rule out that the connection between the SW relic and the tailed radio galaxy is simply due to projection. However, if re-acceleration is invoked, that would alleviate the tight constraints on the acceleration efficiency necessary to reproduce the relic brightness. Moreover, the re-acceleration would also explain the large discrepancy between the X-ray and radio derived Mach number for the SW relic, since the former traces the most recent shock, while the latter reflects the cumulative effect of all events \citep{Kang2021}. Such a scenario would require the main merger-induced shock produced after the core-passage to be followed by other shocks, such as an accretion shock, supporting, for example, a more complex merging scenario involving another infalling subcluster. 

Further, compared to more common arc-like morphologies \citep{Bonafede2009b, DeGasperin2014, Hoang2018, Rajpurohit2024}, where relics are thought to trace outward-traveling shock waves induced in (nearly) head-on mergers, the SW relic features an inverted morphology, bowing away from the cluster, which suggests a more complex merger history. Similar types of inverted morphology are observed in a handful of relics, e.g., in Abell\, 3266 \citep{Riseley2022b}, Coma \citep{Bonafede2022}, and PSZ2\,G145.92-12.53 \citep{Botteon2021}. Recent simulations propose that the ``inverted-morphology" can form when an outward traveling shock wave is bent inward by an infalling galaxy cluster or group \citep{Ludwig2023, Lee2024}, lending further support to a more complex merger.

The lack of an optical counterpart and the relatively flat spectrum rules out a radio-phoenix or AGN relic scenarios for R1, so its nature remains a mystery. A low surface brightness bridge linking the NE relic to R1 indicates a physical connection. One possible explanation is that R1 is associated with the NE relic as relics are known to show diverse morphologies. However, a strong depolarization across R1 and very different rotation measures (although only a small patch is detected) suggests that R1 and NE are separated in Faraday space, i.e., not physically connected. The different trends in the color-color plot across these two structures further support this. R1 may be a radio galaxy. However, we do not find any obvious optical host. There is also a background cluster (RM\,J093941.1+405205.9) to the northwest of R1 located at a redshift of $z=0.50$ \citep{PSZ_Xray}. If R1 is associated with the background cluster, its LLS would be $\sim 600$\,kpc. Therefore, R1 could be a relic associated with the background cluster, located about 1\,Mpc from the cluster center. In conclusion, the nature of R1 is uncertain. 

Finally, the presence of a possible faint, steep-spectrum radio halo coincident with the X-ray ICM serves as further evidence that PSZ2\,G181.06+48.47 is an evolved merger.

\section{Conclusions}
\label{sec:conclusions}

In this paper, we have presented deep, multifrequency, wideband uGMRT and VLA radio observations of the low-mass, merging galaxy cluster PSZ2\,G181.06+48.47. We combined these observations with the published LOFAR data to study the origin of the diffuse emission in the cluster. The cluster was previously classified as a merging system by means of X-ray and optical studies \citep{Botteon2022a}. The two giant diffuse radio sources are located along the merger axis, separated by about 3.5\,Mpc\textemdash the largest known to date.  

Based on our new images, we classify the previously known diffuse radio sources (NE and SW) as radio relics. The NE relic is consistent with an arc-shaped morphology, follows a power-law spectrum, and is strongly polarized at 1--2\,GHz, indicating a weak shock of $\mathcal{M}\sim1.7$. The magnetic field vectors are aligned with the orientation of the radio emission, as expected due to compression by the passage of the merger shock wave. However, the unusually flat spectral index ($\alpha=-0.92\pm0.04$) behavior up to high frequencies and the unrealistic acceleration efficiency requirements are in tension with the DSA model in the test-particle regime. The NE relic can be explained by re-acceleration scenarios. 

The SW relic shows an ``inverted morphology" and exhibits a clear spectral index gradient towards the cluster center. Patches of polarized emission are detected across the SW relic and its integrated radio spectrum is consistent with particle acceleration at the shock of $\mathcal{M}\sim4.8$ by the DSA in the test-particle regime. 

The $\theta\sim 45 \degree$ viewing angle derived from the polarization and radio color-color plot for the NE and SW relic imply that the merger might not be happening in the plane of the sky. Our results suggest that the cluster is in an evolved merger state, likely seen post-apocenter. However, given the current limitations in multi-wavelength data, it is difficult to determine which merger scenario is more likely. If this is the case, the system requires different merging scenarios compared to the edge-on-viewed relics. 

Low resolution uGMRT Band\,3 images reveal significant ($2\sigma$) diffuse emission around the cluster center region (about 1~Mpc). From the non-detection at uGMRT Band\,4, we set an upper limit on the spectral index at $\alpha\leq-2$. The radio power of this emission lies on the known power vs mass relations for radio halos. We classify this emission as a candidate radio halo.

To the north of the NE relic, we find a 370\,kpc (assuming the cluster redshift) diffuse radio source (R1) connecting to the NE relic via a faint bridge-like emission. Its exact nature is unknown. R1 could be a radio galaxy or a radio relic associated with a background cluster. 

Only a limited number of clusters with mass $\leq 5\times 10^{14}M_{\odot}$ are known to host diffuse radio sources. This lower mass range remains largely unexplored due to observational limitations. However, their numbers are increasing thanks to a new generation of telescopes. The discovery of fainter double relics in the low-mass cluster PSZ2\,G181.06+48.47 with the largest relic separation suggests that we are entering into a new territory and that the Square Kilometre Array (SKA) will be a game changer for detecting and understanding the origins of diffuse radio sources in low-mass clusters. 

\section*{acknowledgments}
We thank Hiroki Akamatsu and Grant Tremblay for useful discussions. A. Stroe gratefully acknowledges the support of a Clay Fellowship and the NASA 80NSSC21K0822 and Chandra GO0-21122X grants. M. J. Jee acknowledges support for the current research from the National Research Foundation (NRF) of Korea under the programs 2022R1A2C1003130 and RS-2023-00219959. L. Lovisari acknowledges the financial contribution from the INAF grant 1.05.12.04.01. This work is based on observations from GMRT, which is run by the National Centre for Radio Astrophysics (NCRA) of the Tata Institute of Fundamental Research (TIFR). This research has made use of VLA observations, which is operated by the National Radio Astronomy Observatory, a facility of the National Science Foundation operated under cooperative agreement by Associated Universities. This work used LOFAR observations, which was designed and constructed by ASTRON. This work has also made use of observations with the Chandra X-ray Observatory and \xmm, an ESA science mission with instruments and contributions directly funded by ESA Member States and the US (NASA).

\vspace{5mm}
\facilities{VLA, GMRT, LOFAR, XMM, CXO, SDSS, PanSTARRS}

\software{
Astropy \citep{astropy2013, astropy2018},
APLpy \citep{aplpy},
Matplotlib \citep{matplotlib},
TOPCAT \citep{topcat},
DS9 \citep{ds9},
CASA \citep{casa2022}}

\bibliography{PSZ_radio}{}
\bibliographystyle{aasjournal}

\end{document}